\documentclass[10pt,preprint]{aastex}

\shorttitle{Star-Forming Torus and Black Hole Mass in NGC\,3227}
\shortauthors{Davies et al.}

\newcommand{\kms}{\,\hbox{\hbox{km}\,\hbox{s}$^{-1}$}}

\begin{document}

\title{The Star-Forming Torus and Stellar Dynamical Black Hole Mass in
  the Seyfert~1 Nucleus of NGC\,3227\footnote{Based on observations at
    the European Southern Observatory VLT (074.B-9012).}}

\author{R.I. Davies$^1$, J. Thomas$^{1,2}$, R. Genzel$^{1,3}$, 
  F. Mueller S\'anchez$^1$, L.J. Tacconi$^1$, A. Sternberg$^4$, 
  F. Eisenhauer$^1$, R. Abuter$^1$, R. Saglia$^{1,2}$, R. Bender$^{1,2}$}
\affil{$^1$ Max-Planck-Institut f\"ur extraterrestrische Physik, 
Postfach 1312, 85741, Garching, Germany}
\affil{$^2$ Universit\"ats-Sternwarte, Scheinerstrasse 1, 81679,
  M\"unchen, Germany}
\affil{$^3$ Department of Physics, University of California at
  Berkeley, 366 Le Conte Hall, Berkeley, CA 94720-7300, USA}
\affil{$^4$ School of Physics and Astronomy, Tel Aviv University, Tel
  Aviv 69978, Israel}



\begin{abstract}

We report $R\sim4300$ Very Large Telescope SINFONI adaptive optics
integral field K-band spectroscopy of the nucleus of the Seyfert~1
galaxy NGC\,3227 at a spatial resolution of 0.085\arcsec\ (7\,pc).
We present the morphologies and kinematics of emission lines and
absorption features, and give the first derivation of a black hole
mass in a Seyfert 1 nucleus from spatially resolved stellar dynamics.
We show that the gas in the nucleus has a
mean column density of order $10^{24}$--$10^{25}$\,cm$^{-2}$ and that
it is geometrically thick, in agreement with the standard `molecular
torus' scenario.
We discuss which heating processes may be
responsible for maintaining the vertical height of the torus.
We have also resolved the nuclear stellar distribution, and find that 
within a few parsecs of the AGN there has been an intense
starburst. 
The most recent episode of which began $\sim$40\,Myr
ago but has now ceased.
The current luminosity of stars within 30\,pc of the AGN,
$\sim$$3\times10^9$\,$L_\odot$, is comparable to that of the AGN.
Based on a comparison of the respective size scales, we argue that the
star formation has been occuring in the obscuring torus.
Finally, we present the first derivation of a black hole mass in a
Seyfert 1 nucleus from stellar dynamics which marginally spatially resolve
the black hole's sphere of influence.
We apply Schwarzschild orbit superposition models to our
full 2-dimensional data and derive the mass of
the black hole, paying careful attention to the input parameters which
are often uncertain:
the contribution of the large scale bulge and its mass-to-light ratio;
the recent star formation in the nucleus and its mass-to-light ratio;
the contribution of the gas mass to the potential;
and the inclination.
Our models yield a 1$\sigma$ range for the black hole mass of 
M$_{\rm BH} = 7\times10^6$--$2\times10^7$\,$M_\odot$.

\end{abstract}


\section{Introduction}
\label{sec:intro}

Three issues concerning active galactic nuclei (AGN) which
are currently receiving considerable attention, and which we address
in this paper, are: 
the relationship of the AGN to the surrounding star formation, 
the central black hole mass, 
and the physical properties of the obscuring torus.

The issue of star formation in AGN has long been contentious,
although it is now clear that on 0.1--1\,kpc scales star
formation is an important process in all types of AGN \citep{cid04}.
In particular empirical population synthesis of Seyfert~2 nuclei
\citep{cid01,gon01,sto01} reveals evidence for recent ($<$1\,Gyr) star
formation within a few hundred 
parsecs of the nuclei in about 40\% of Seyfert 2 nuclei.
While some authors \citep[e.g.][]{jog01,gu01} reach similar
conclusions, not all concur \citep{iva00}.
Mid- and Far- infrared ISO data \citep{ver05} also have much to offer
on this issue.
The situation regarding Seyfert 1 nuclei is rather more uncertain, and a
criticism (equally applicable to Seyfert~2s) often levelled is that 
it lies so far from the AGN that it cannot be
associated with or strongly influencing the characteristics of the AGN
\citep{hec97}.
Nevertheless, \cite{dav04a,dav04b} report two cases where powerful
starbursts do exist within the central few tens of parsecs of type~1
AGN.

Beyond this debate is the fundamental question concerning the physics
linking the two phenomena through a causal relationship
\citep{shl89,goo04,thomp05} -- which is related to, although more
complex than, driving gas to small scales.
While observation and theory provide strong indications linking gas
inflow and circumnuclear starbursts to AGN and nuclear starbursts, the 
statistical evidence is tenous \citep{kna04,wad04,shl03}.
For example, HST imaging shows no dependence of AGN activity on
the presence of a nuclear spiral \citep{mart03}.
Similarly, AGN are equally common in galaxies with and
without a compact nuclear gas component \citep{gar03}.

A parallel theme is the molecular obscuring torus \citep{ant85,ant93},
the heart of the unification scheme.
Many models of the thermally re-radiated spectrum
of the putative torus have been constructed, both as a uniform ring
\citep{pie92a,pie93,gra94,sch05}, and also as a clumpy medium
\citep{nen02} as was proposed for NGC\,1068 a decade earlier by
\cite{cam93}.
In all cases, the crucial properties are a spatial scale of tens (to
perhaps hundreds) of parsecs, a vertically extended geometry, and a
high column density.
In the last few years, models have begun to consider
star formation within the torus \citep{wad02,thomp05}, giving credence
-- albeit with a different perspective -- to some early ideas about
stellar light in AGN spectra.

The mass of the central supermassive black hole $M_{\rm BH}$ is the
third issue we address.
The masses of black holes in nearby AGN are most commonly estimated by
reverberation mapping \citep{onk04}, a technique that can be
extended to higher redshift via additional scaling relations
\citep{kas00,kas05,ves02,ves04}.
Providing an independent measure of $M_{\rm BH}$ for reverberation
masses would allow one to begin to understand the geometry of
the broad line region, which is the limiting factor in the accuracy of
the method \citep{hor04}.

Estimates of such masses are important with respect to the relation
between the $M_{\rm BH}$ in the center of a stellar spheroid and the
velocity dispersion $\sigma_*$ \citep{fer00,geb00}.
This relation has superseded similar ones using the luminosity or mass
of the spheroid \citep{kor95,mag98} to become a cornerstone in the
cosmological context of galaxy evolution and black hole growth.
It is generally accepted that the $M_{\rm BH}-\sigma_*$ relation should be
valid for all spheroids irrespective of whether the black hole is
quiescent or active, whether the spheroid in which it lies is a
globular cluster or a giant elliptical galaxy, whether the spheroid is
embedded in a gaseous disk or not, and so on. 
However, almost without exception the black hole masses that are
considered `reliable' -- typically those based on stellar
kinematics and for which the radius of influence of the black hole has
been resolved -- have been derived only for nearby bulge
dominated E/S0 galaxies \citep{tre02,mar03,fer05}.
The smaller bulges of spiral galaxies imply lower $M_{\rm BH}$, making
it difficult to spatially resolve the stellar kinematics.
And for AGN, where the black hole is by definition active rather than
quiescent, the glare of the AGN itself is an added obstruction.
To date there is only one such published mass,
$\sim2\times10^8$\,M$_\odot$ for Cen\,A \citep{sil05}.
Despite this galaxy's proximity, seeing limited
observations are only able to resolve the 
radius of influence of the black hole because the black hole 
is at least 5 times more massive than expected from the 
$M_{\rm BH}-\sigma_*$ relation.
Given this result, it is crucial that black hole masses are
derived using stellar kinematics at high spatial resolution for more
AGN.

In this paper we look in detail at these topics for the Seyfert~1
nucleus of NGC\,3227.
We probe the star formation and molecular gas in the central 80\,pc at
a spatial resolution of only a few parsecs.
In particular we discuss how the young stars and gas relate to 
our understanding of the canonical torus.
In addition we derive the mass of the black hole from Schwarzschild
modelling of the stellar dynamics.
This is the first time that this has been possible for a Seyfert~1
galaxy, where using integral field capability we measure the
kinematics across the full 2-D field.
We pay careful attention to issues which are often rather uncertain in
Schwarzschild orbit superposition models:
the inclination of the system, the mass-to-light ratio, and the
contribution of gas to the gravitational potential.

The distance to NGC\,3227 of 17\,Mpc (for H$_0=70$--75\kms\,Mpc$^{-1}$) is
estimated from the 1250\kms\ luminosity weighted average velocity of
the 13 galaxies in the group of which it is a member \citep{gar93}.
Its infrared (8--1000\micron) luminosity is then
$\log{L_{\rm IR}/L_\odot}=9.93$.
However, this does not represent the bolometric luminosity since the
spectral energy distribution $\lambda F_\lambda$ is approximately flat
in the range 0.3--1000\micron, and perhaps to even shorter wavelengths.
Taking this into account we estimate the bolometric
luminosity to be $\log{L_{\rm bol}/L_\odot}=10.2$. 

\section{Observations and Data Reduction}
\label{sec:obs}

The data presented here were obtained on 21~Dec~2004 using SINFONI
\citep{eis03a,bon04} on the VLT UT4.
The instrument consists of a cryogenic near infrared integral field
spectrometer SPIFFI \citep{eis03a,eis03b} coupled to a visible curvature
adaptive optics (AO) system \citep{bon03}.
The AO module was able to correct on the nucleus of NGC\,3227 (for
which it measured R=13.9\,mag) in
seeing of $\sim$0.6\arcsec, to reach nearly the diffraction limit of
the telescope in the K-band (an estimated 15\% Strehl).
With the appropriate pixel scale selected, the spectrograph was able,
in a single shot, to obtain spectra covering the whole of the K-band
(approximately 1.95--2.45\micron) at a spectral resolution of
$R\sim4300$ for each $0.0125\arcsec\times0.025\arcsec$ pixel in a
$0.80\arcsec\times0.80\arcsec$ field of view.
A total of 3 sky and 6 on-source exposures of 600\,sec each, dithered
by up to 0.2\arcsec, were combined to make the final data cube.

The data were reduced using the SINFONI custom reduction package
SPRED.
This performs all the usual steps needed to reduce near infrared
spectra, but with additional routines for reconstructing the data
cube.
Because the sky airglow did not vary much during the observations
and is low compared to the read noise at such small pixel scales, it
was possible to make a combined sky frame which 
could be subtracted from all the on-source frames without leaving OH
line residuals.
Following this step, the frames were flatfielded and corrected for
dead/hot pixels.
The data were then interpolated to linear wavelength and spatial
scales, after which the slitlets were aligned and stacked up to
create a cube.
Finally the atmospheric absorption was compensated using the B9 star
HD\,83434;
and flux calibration was performed using HD\,83434 (K=6.917) and
HD\,87015 (K=6.144), which both yielded a zeropoint of 16.08\,mag.

\paragraph*{Spatial Resolution}

\begin{figure}
\epsscale{0.8}
\plotone{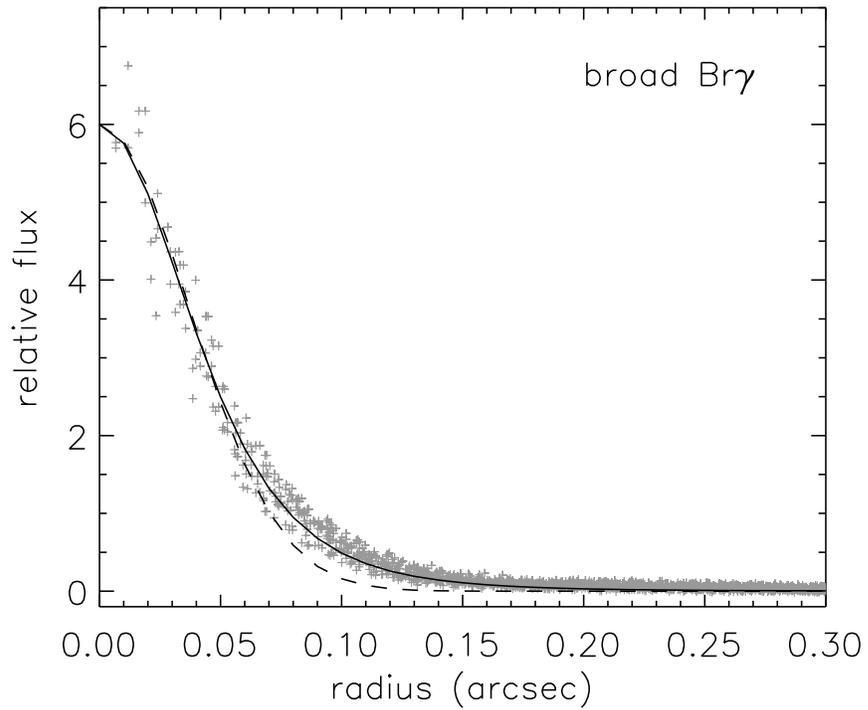}
\caption{Azimuthally averaged profile of the broad Br$\gamma$ line
  map, used to estimate the spatial resolution. The data points
  (values for every pixel within 0.3\arcsec\ of the centre) are
  marked as grey crosses and show no asymmetric artifacts
  which could result from the AO correction. Overdrawn is a Gaussian
  profile (dashed 
  line) and a Moffat profile (solid line). Although the latter better
  reproduces the wings in the profile, both appear to be reasonable
  apprxoimations to the profile and both yield the same FWHm
  resolution of 85\,mas.}
\label{fig:brg_res}
\epsscale{1.0}
\end{figure}

No additional point-spread function (PSF) calibration frames using
stars were taken.
This is primarily because, although in principle one can match the
brightness of a calibration star on the wavefront sensor to the AGN, it
is not possible to replicate either the spatial extent of the AGN or
the background galaxy light associated with it.
Together with changes in the ambient seeing which can occur on
relatively fast timescales, this can result in a
potentially considerable mismatch between the science and calibration
PSFs \citep{dav04c}.
Since, for the analysis presented here, a highly accurate PSF is not needed,
we have instead made use of the broad
Br$\gamma$ emission which is unresolved: it has been measured to be of
order 0.02\,pc by \cite{sal94} and \cite{onk03}.
This has the advantage of providing the resolution directly from the
science frames, and includes all effects associated with shifting and
combining the cube, and also smoothing (a
$0.0375\arcsec\times0.0375\arcsec$ median filter was applied to the
final cube).
The PSF we use for convolving our models is then approximated by a
symmetrical fit to the broad Br$\gamma$ emission as shown in
Fig.~\ref{fig:brg_res}.
In this figure we have plotted the value for every pixel within
0.3\arcsec\ of the centre, from which it can be seen there are no
asymmetric artifacts 
resulting from the AO correction.
Overdrawn are both Gaussian and Moffat profiles. 
The latter, as expected, matches the faint extended wings in the profile;
but both reproduce the core of the profile equally well.
They yield a FWHM resolution of 0.085\arcsec.

\paragraph*{Derivation of the Kinematics and their Uncertainties}

The velocity and dispersion of the emission lines is found by
simultaneously fitting the continuum level with a linear function and
the line itself with a Gaussian.
The uncertainty of the fit is estimated using Monte Carlo techniques
assuming that the noise is uncorrelated and the line is well
represented by a Gaussian. 
The method involves adding a Gaussian with the measured properties to
a spectral segment with the same noise statistics as the data, and
re-fitting the line to yield a new set of Gaussian parameters. 
After repeating this 100 times, the standard deviation of the center
and dispersion are used as the uncertainties for the velocity and line
width.

The shape of the CO absorption bandheads in a galaxy spectrum arises
from the intrinsic spectral profile of the stars
themselves convolved with a broadening profile which carries
the information about the kinematics.
The use of template stars (we chose the K0\,I star HD\,179323) allows
one to separate these two components.
One technique commonly used is deconvolution with the Fourier
cross-correlation quotient method.
The principal advantage of this method is that it allows one to
recover the full broadening profile even if it is not well represented
by a function such as a Gaussian (plus Hermite terms).
However, because the Fourier transform of the object is divided by
that of the template, the noise is amplified;
and while it can be reduced by Wiener filtering, setting the
optimal parameters of the filter is only possible if the noise and
signal can be properly distinguished in Fourier space.
If this is not done correctly, it can be difficult to obtain the
correct object dispersion.
The alternative is to convolve the template with an analytical
function and iteratively minimise its difference to the object using a
$\chi^2$ criterion.
The limitation of this technique is its speed and the fact that the
convolution function needs to be relatively simple.
On the other hand, it is less adversely affected by noisy data and one
can easily reject bad data values.
In addition it provides a clear path to estimating the uncertainties
using standard techniques.
One can estimate the confidence level for each parameter separately
by finding the value for which, with all the other parameters
re-optimised, $\chi^2$ increases by 1.

It is the latter method which we have used because the individual
spectra for each spatial pixel are rather noisy,
and we only extract the usual three parameters for a Gaussian fit.
Tests have shown that for these data it is not possible to determine
coefficients for the Hermite polynomials which would indicate the
deviations from a Gaussian.

We have only made use of the first (i.e. CO\,2-0) bandhead, because 
[Ca\,{\sc viii}] emission at the edge of the CO\,3-1 strongly biasses
  the derived velocities (making them much more positive where the 
  line emission is strongest) when this second bandhead is used.
The third CO\,4-2 bandhead adds very little because it is strongly
affected by residual atmospheric features.

\paragraph*{Smoothing}

All the images presented in this paper have been smoothed using
optimal Voronoi tessellations, as described and implemented by \cite{cap03}.
This scheme uses adaptive binning to group pixels so that
the combined signal to noise of each group is above a minimum
threshold. 
Pixels which are already above the threshold are not binned.
Thus the final resolution varies across the image, although at no
point is it improved beyond the original resolution in the data.
 
The algorithm bins pixels together into groups by accreting new pixels
to each group according to how close they are to the centroid of the
current group. 
Checks ensure that each time a pixel is added to a group, both
the group remains `round' and the signal to noise increases.
The resulting groups then provide a set of positions (centroids) and
mean fluxes which are used as the initial `generators'.
A further algorithm optimises the generator
configuration based on a centroidal Voronoi tessellation.
The final set of generators are the positions of the flux-weighted
centroids of each binned group, and have the property that each pixel
in the original image is assigned to the group
corresponding to the nearest generator.
Thus the positions of the generators provide all the information
necessary to re-create the binned image.

Since it can be hard for the human eye to comprehend the binned images
created in this way, we have -- for the purposes of presentation only --
applied a final step of interpolating all 
the pixel values from the generators of the Voronoi tessellation using
a minimum curvature spline surface (as implemented in IDL version 6.0).
In the resulting smooth images, pixels with high signal to noise
(i.e. unbinned pixels) have very nearly the same value as the original
image.

\section{The Circumnuclear Stellar Ring}
\label{sec:ring}

When the pixel scale of SINFONI is matched to the diffraction limit of
the VLT, its field of view is only $\sim$1\arcsec.
An important step is therefore to understand the larger scale context
in which these data reside.
To facilitate this we have used H-band data -- which trace the
stellar light with less bias to recent star formation than
optical data -- from both the 2MASS large galaxy atlas
\citep{jar03} on scales of 2--100\arcsec, and the Hubble Space
Telescope ({\it HST}) archive on scales of 0.2--10\arcsec\ (proposal
7172, Rieke).
Before analysing the HST F160W image, it was deconvolved using a PSF
generated by Tiny Tim 6.2, rotated to the standard orientation and trimmed
to 13.5\arcsec\ on a side.
Isophotal fitting (using the {\em ellipse} task in {\sc iraf}) was
performed on both these images and the resulting profiles scaled
according to their overlapping region.
Additionally, a bulge-plus-disk model (the parameters of the disk
being fixed at the values determined from the large scale 2MASS data)
was fit to the F160W image, excluding radii smaller than 1.6\arcsec,
using a $\chi^2$ minimisation.
This model was then divided into the F160W image in order to
investigate the excess continuum emission apparent at radii of
1--1.5\arcsec.

On the largest scales there is an almost perfect exponential profile
with a disk scale length of 26.8\arcsec\ (2.1\,kpc) and an axis ratio
of 0.4--0.6 (corresponding to an inclination of 55--65$^\circ$) at a
position angle (PA) of $\sim-30^\circ$.
At smaller scales, there is a smooth transition to an
$r^{1/4}$ profile with an effective radius of 
$r_{\rm eff} = 3.4$\arcsec\ (270\,pc), which dominates the emission at
radii 2--9\arcsec.
This $r_{\rm eff}$ is rather larger than the 2.6\arcsec\ found by
\cite{nel04} from optical data, probably because we have less bias
from nuclear star formation and the AGN.

\begin{figure}
\epsscale{0.8}
\plotone{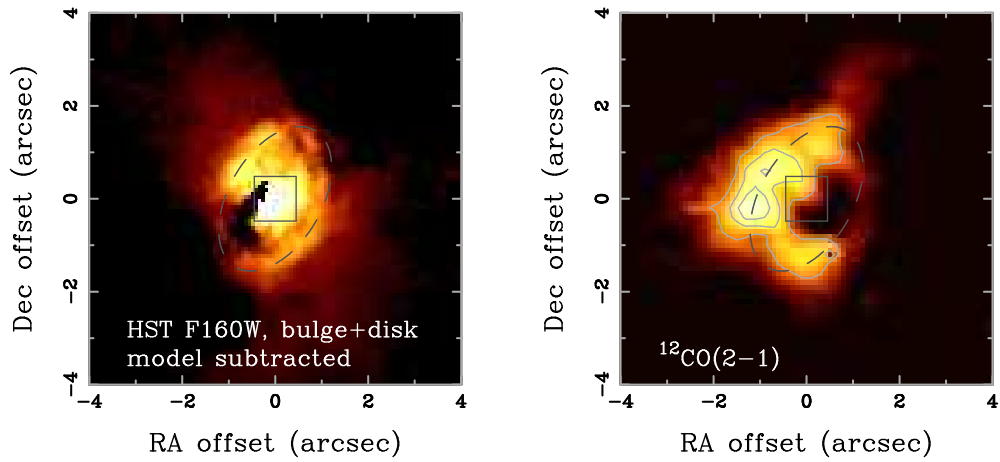}
\caption{Left: the HST F160W (square-root scaling) image after subtracting the
  bulge-plus-disk model derived from the 2MASS image and the HST image
  at $r>1.6$\arcsec.
The field of view of SINFONI is shown as a box in the centre.
The outline of the stellar ring is drawn as a dashed ellipse, with axis
  ratio 0.6 at position angle $-30^\circ$.
Right: CO\,(2-1) molecular gas map from \cite{sch00}, with contours at
  40, 60, and 80\% of the peak.
The same box and ellipse as at left are also marked on this
  image.
North is up and east is left.
}
\label{fig:hst_div}
\epsscale{1.0}
\end{figure}

What was previously reported by \cite{cha00} as a possible knotty
one-armed spiral is now revealed in  Fig.~\ref{fig:hst_div} as a ring
of excess continuum with radius 1.7\arcsec\ (140\,pc).
Its axis ratio of 0.6 and PA of $-30^\circ$) are consistent with both
the large scale disk and
also the circumnuclear molecular gas ring reported by both
\cite{sch00} and \cite{bak00}.
The kinematics of the CO\,(2-1) indicate that the gas lies in a disk;
and the stellar ring indicates that the stars also lie
in a disk, since resonances such as these are features of disk
dynamics.
The similarity of the loci traced in both cases suggests that the
stellar and gas disks are in fact the same.
The ring was hypothesised by \cite{sch00} to be at the location of an
inner Lindblad resonance (ILR) associated with a secondary inner bar on
scales smaller than 20\arcsec, which is itself the ILR for the primary bar.
Evidence for such a secondary bar was presented by \cite{bak00} who noted
elongated molecular structures along the major axis at radii of
10--15\arcsec, which he suggests may be concentrations of gas at the
leading edges of a bar.
However, he cautions that the velocities are different to those
expected for gas inflow along the bar.

\begin{figure}
\epsscale{0.9}
\plotone{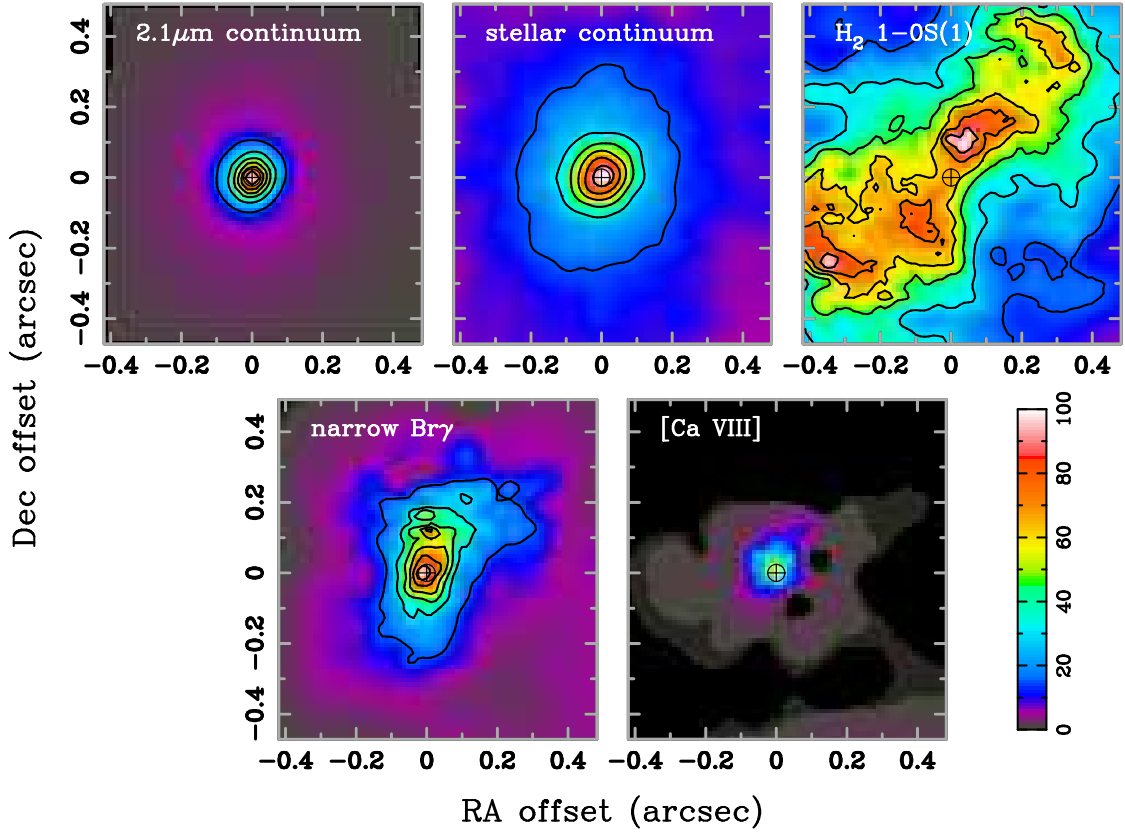}
\caption{Images from the SINFONI data cube showing the primary
  continuum and emission line
  morphologies in the central arcsec of NGC\,3227. In each case, the
  colour scale stretches from 0--100\% of the peak flux, and contours
  are spaced equally between 15\% and 90\% of the peak flux.
A crossed circle indicates in each panel the peak of the continuum
  emission.
The maps show, from left to right and top to bottom: 
2.1\micron\ continuum, stellar continuum (derived from the stellar
  absorption), H$_2$ 1-0\,S(1), narrow Br$\gamma$, and [Ca\,{\sc viii}].
North is up and east is left.
}
\label{fig:fluxmaps}
\epsscale{1.0}
\end{figure}

\begin{figure}
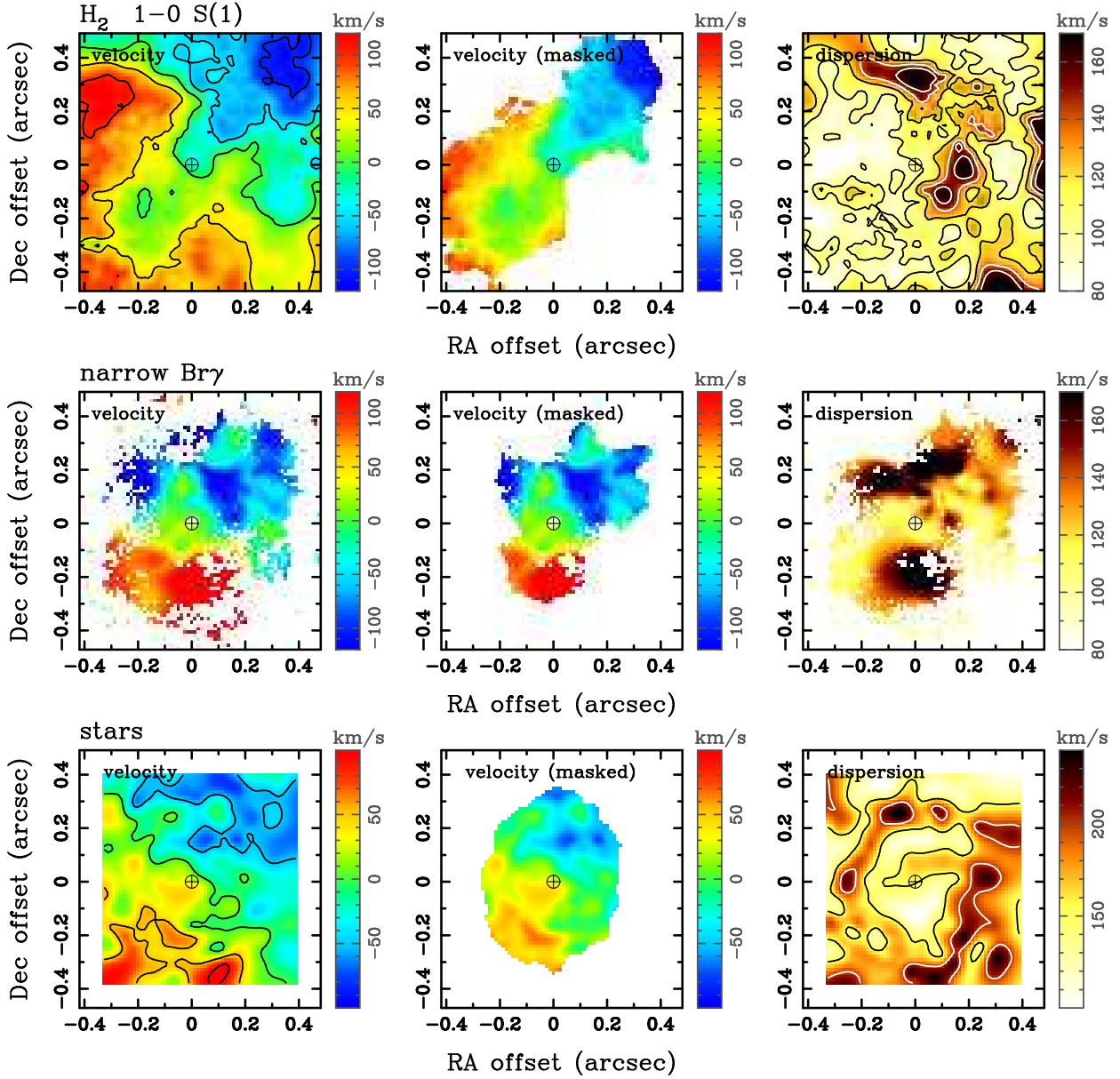

\epsscale{1.0}
\plotone{f4a.eps}
\plotone{f4b.eps}
\plotone{f4c.eps}
\caption{Measured kinematics for the gas and stars in the central
  arcsec of NGC\,3227.
Top: H$_2$ 1-0\,S(1) line; 
middle: narrow component of the Br$\gamma$ line;
bottom: stars.
Left: velocity field, determined as described in Section~\ref{sec:obs};
Centre: the same velocity field, masked to include only those pixels
with higher fluxes -- in each case, these pixels make up 2/3 of the total
flux;
Right: dispersion.
The maps for emission lines are at a resolution of 0.085\arcsec; those
  for the stars have had additional smoothing, and are at an effective
  resolution of 0.115\arcsec.
Note that in the line dispersion maps, the data have not been
  corrected for the instrumental resolution, which corresponds to
  $\sigma = 30$\kms.
}
\label{fig:velfield}
\epsscale{1.0}
\end{figure}

The 1\arcsec\ field of view of SINFONI lies entirely inside the
140\,pc ring, as indicated on Fig.~\ref{fig:hst_div}.
Its view is presented in Fig.~\ref{fig:fluxmaps}, where
we show images of the 2.1\micron\ continuum,
the stellar continuum measured from the CO\,2-0 absorption bandhead,
the 2.12\micron\ H$_2$ 1-0\,S(1) line, 
the 2.17\micron\ Br$\gamma$ line, 
and the coronal [Ca\,{\sc viii}] line at 2.32\micron.
Fluxes and flux densities are summarised in Table~\ref{tab:flux}.
The corresponding kinematics for the 1-0\,S(1) and Br$\gamma$ lines
and the stellar CO\,2-0 absorption are shown in
Fig.~\ref{fig:velfield}.
In the lefthand panels we show the full field that can be measured.
In the centre panels we have applied a mask to show only velocities
associated with the brightest pixels: those that contribute 2/3 of the
total flux, and hence the bulk of the emitting medium, in the field.
Velocity dispersions are shown in the righthand panels.

\section{Nuclear Star Formation}
\label{sec:nuc_sf}

There is some evidence in the literature that points to a scenario involving
recent vigorous star formation in the nucleus of NGC\,3227.
Based on an analysis of seeing limited near infrared integral field
spectroscopy, \cite{sch01} argued that there was a 25--50\,Myr cluster
present, although they could not rule out a much older population.
In addition, there is the detection of the 3.3\micron\ polycyclic
aromatic hydrocarbon (PAH) feature in a
0.8\arcsec$\times$1.7\arcsec\ slit aperture by \cite{rod03}.
They reported that the ratio $L_{3.3}/L_{\rm IR}\sim7\times10^{-5}$ is lower
than the mean for starbursts but typical of Seyfert 2s.
PAHs have also been detected in AGN where star formation is also
active, perhaps most notably Mkn\,231 \citep{rig99} and NGC\,7469
\citep{maz94}; 
\cite{dav04a,dav04b} have shown beyond doubt that these galaxies host
massive nuclear star formation.
In this Section we address the questions of the presence and age of a
distinct stellar component in the nucleus.
We consider evidence from the stellar absorption features
(Sections~\ref{sec:EW} and~\ref{sec:star_prof}), 
the narrow Br$\gamma$ (Section~\ref{sec:brg}),
and radio continuum (Section~\ref{sec:radcont}).
Finally we apply population synthesis models in Section~\ref{sec:stars}.

\subsection{Late Type Stars and Stellar Luminosity}
\label{sec:EW}

\begin{figure}
\epsscale{0.6}
\plotone{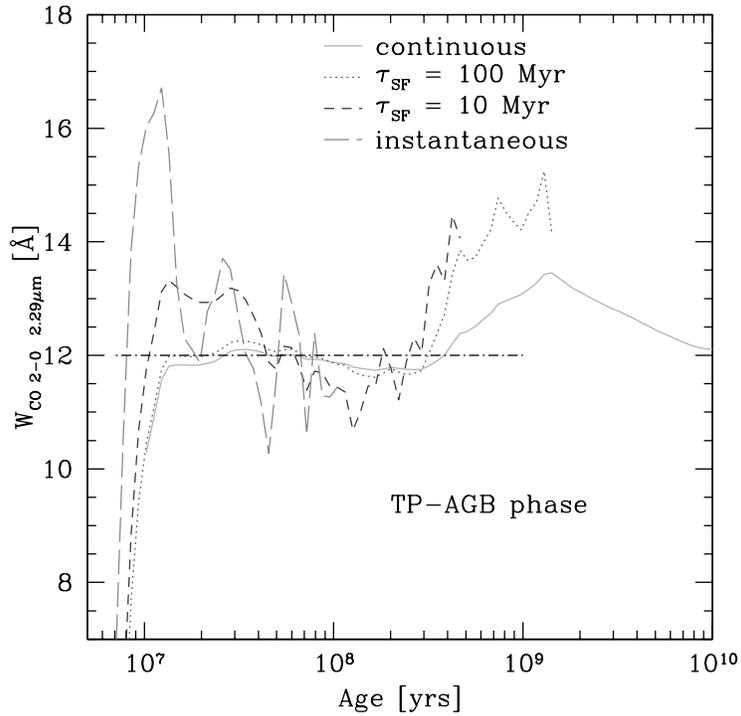}
\caption{Equivalent width of the 2.294\,\micron\ CO\,2-0 bandhead as a function
  of age, calculated using STARS (which includes the thermally pulsing
  asymptotic giant branch stars). 
  Several different star formation histories
  are shown: instantaneous, a decay time
  of 10\,Myr, a decay time of 100\,Myr, and continuous. For each history, data
  are plotted where the K-band luminosity is at least 1/15 of its
  maximum. This indicates that one expects 
$W_{\rm CO}\sim12$\,\AA\ for nearly all star forming scenarios.
}
\label{fig:stars}
\epsscale{1.0}
\end{figure}

The CO bandheads at $\lambda>2.29$\micron\ are often used as tracers of
star formation in the K-band since, once they first appear, late type
stars dominate the near infrared stellar continuum.
Under the assumption that either the equivalent width of the CO\,2-0
bandhead, $W_{\rm CO}$, is independent of stellar population or that
the stellar population does not change across the region of interest,
the absolute absorption in the bandhead provides a direct tracer of
stellar luminosity.
However, given that \cite{for00} and others have shown that 
$W_{\rm CO}$ in individual stars can vary from 0--20\AA\ depending on
stellar type (effective temperature), the validity of this assumption
is far from clear.
Fortunately, there is observational and theoretical evidence that it
is valid.
\cite{oli95} found very little variation in $W_{\rm CO}$ between
elliptical, spiral, and H{\sc ii} galaxies. 
In addition, star cluster models indicate that for ensembles of stars,
$W_{\rm CO}$ is much more stable, reaching a value close to 12\AA\ once the
age of the cluster has exceeded 10\,Myr.
This is demonstrated in Fig.~\ref{fig:stars} which shows how 
$W_{\rm CO}$ varies as a function of age for various star formation
histories.
The data were generated using the population synthesis code STARS
\citep{ste98,tho00,dav03,ste03,dav05},  
which calculates the distributions of stars in the
Hertzsprung-Russell diagram as a function of age for exponentially
decaying star formation rates.
Using empirically determined $W_{\rm CO}$ from library spectra
\citep{for00}, the code then computes the 
time-dependent $W_{\rm CO}$ for the entire cluster of stars.
The code includes the thermally pulsing asymptotic giant branch
(TP-AGB) stars which have a very significant impact on the 
depth of the absorption features at ages of 0.4-2\,Gyr
\citep{for03,mar05}.
Throughout this work, we adopt a solar metallicity Salpeter IMF in the
range 1--100\,M$_\odot$. 

Fig.~\ref{fig:stars} shows that once CO absorption is present,
$W_{\rm CO}$ does not deviate from 12\AA\ by more than 20\% except in
the special case of instantaneous star formation with an age of less
than 15\,Myr.
This insenitivity of $W_{\rm CO}$ to star formation history means
that, although it cannot differentiate between young and old stellar
populations,  it can be used to distinguish the stellar and non-stellar
continuum and to trace the total stellar luminosity profile.
The observed $W_{\rm CO}=3.6$\,\AA\ (measured in the bandpass
prescribed by \citealt{for00}) implies a mean dilution (i.e. ratio of
total continuum to stellar continuum) in a 0.8\arcsec\ aperture of 3.3
at 2.3\micron.
The average dilution over the whole K-band will be less -- as
small as 2.3 if the hottest dust associated with the AGN is at a
typical temperature of 500\,K within this aperture (or hotter but
also reddened).
We have adopted an intermediate value of 2.8 corresponding to a
characteristic temperature of 1000\,K.
Within the aperture, we find a K-band magnitude of 10.28 (a little
brighter than that given by \citealt{sch01}, perhaps due to their
lower resolution or because of the wings often associated with
the PSF of shift-and-add data), which implies a stellar 
magnitude of 11.4 or equivalently a 1.9--2.5\micron\ luminosity of 
$\log{L_{\rm K}/L_\odot} = 7.8$. 

This luminosity includes both a possibly young population as well as
the old bulge population.
Fortunately, the respective contributions from these two components
can be disentangled, as the top panel of Fig.~\ref{fig:prof} shows.
This figure includes data at different scales from various sources,
all of which have pros and cons:
2MASS H-band covers the largest scales but cannot probe scales less
than 2--3\arcsec\ due to limited resolution;
HST F160W image covers intermediate and small scales at high
resolution, but cannot probe radii smaller than $\sim$0.2\arcsec\ due
to the bright point source associated with the AGN;
SINFONI total K-band continuum covers the central arcsec at high
spatial resolution, but is limited by the AGN at radii less than
$\sim$0.05\arcsec;
SINFONI stellar K-band continuum probes the very smallest scales
without being affected by the AGN, but is noisier (see
Fig.~\ref{fig:fluxmaps}).
The large scale disk and bulge models discussed in
Section~\ref{sec:ring} reveal an excess of emission at radii
less than 0.5\arcsec\ that increases to become very significant closer
than 0.2\arcsec\ from the centre.
Presumably this excess is associated with the recent star formation that
is also the origin of the narrow Br$\gamma$ flux.
It accounts for 60\% of the continuum in a 0.8\arcsec\ aperture.
Hence the luminosity of the young stellar population is 
$\log{L_{\rm K}/L_\odot} = 7.6$. 

When compared to the luminosities of star clusters predicted by STARS
(e.g. see \citealt{dav03}) or to individual clusters in other nearby
galaxies, this suggests a very significant (probably young) stellar
component within 30\,pc of the AGN.

\subsection{Nuclear Stellar Luminosity Profile}
\label{sec:star_prof}

The 2.1\micron\ continuum (similar to a broadband K image) is
dominated by the non-stellar continuum associated with the AGN and is
barely resolved, with a FWHM of 0.10\arcsec\ (cf the resolution
of 0.085\arcsec).
On the other hand, 
a simple size measurement of the stellar continuum yields a FWHM of
0.17\arcsec\ indicating that the nuclear stellar component is easily
resolved, having an intrinsic size scale of $\sim$12\,pc.

\begin{figure}
\epsscale{0.6}
\plotone{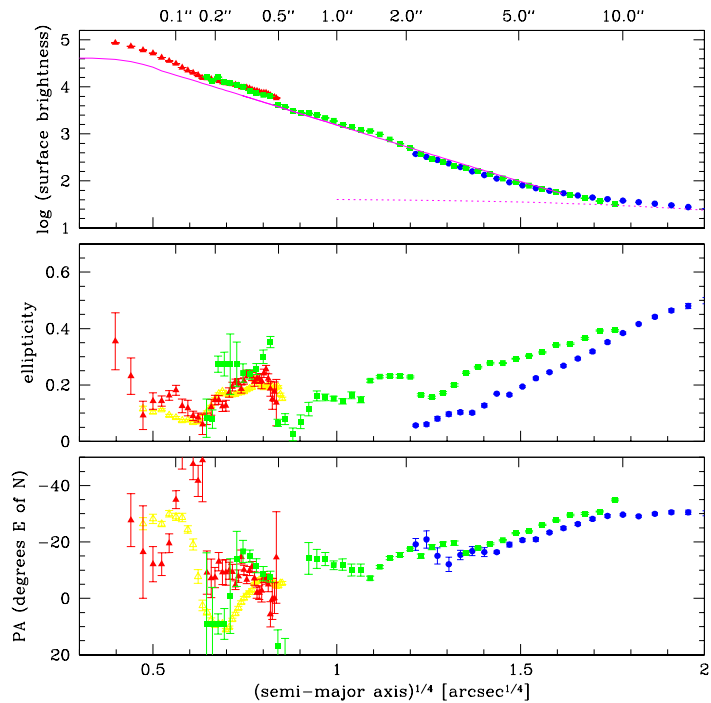}
\caption{Luminosity profile (top), ellipticity (centre), and position
  angle (bottom) as functions of radius for NGC\,3227.
Data from different sources are denoted by:
blue circles for 2MASS H-band, green squares for HST F160W, red
  triangles for SINFONI stellar K-band continuum, yellow triangles for
  SINFONI total K-band continuum.
In the top panel, the relative scaling has been adjusted according to
  the overlapping regions.
The dotted line indicates the large scale disk; the solid line the
  bulge plus disk. A clear emission excess is apparent in the central
  arcsec, associated with changes in ellipticity and position angle.
}
\label{fig:prof}
\epsscale{1.0}
\end{figure}

The axis ratio of 0.8 and PA $-10^\circ$ evident in
the outer isophotes of the total stellar distribution -- also apparent
in the faintest levels 
of the 2.1\micron\ image -- differ only slightly from those of the
140\,pc scale ring.
This can be seen in Fig.~\ref{fig:prof}:
from radii of 10\arcsec\ where the bulge begins to dominate down to a
scale of 2\arcsec, the ellipticity of the isophotes decreases;
but, as measured in the HST F160W data as well as the SINFONI
continuum and stellar bandhead data,
it increases again briefly at both the 1.7\arcsec\ radius of the ring and
also at radii 0.2--0.5\arcsec.
The stellar kinematics (Fig.~\ref{fig:velfield}) show ordered
rotation, albeit with a very large 
velocity dispersion, at PA $-30^\circ$ to $-45^\circ$.

It may be, as appears to be the case in NGC\,7469
\citep{dav04a}, that the nuclear stellar cluster is triaxial,
leading to an offset between the major (isophotal) axis and the
kinematic axis.
A triaxial potential could conceivably arise in a situation where a
nested bar -- perhaps associated with the 140\,pc ring -- is
dynamically heated and forms a mini pseudobulge \citep{kor04}.
The timescale for such heating is of the order of 10$^9$\,yr for
kpc-scale bulges and presumably shorter for nested bars where the
pattern speed is faster and significant energy may be injected either
by star formation or the AGN.
If the heating is due to buckling instability, then it will result in
the bar becoming weaker and more centrally compact \citep{rah91}, at
least qualitatively consistent with what we are seeing here.

However, in interpreting the stellar data one
needs to bear in mind that it comprises similar contributions from the
bulge and the nuclear component, the observed properties being the
combination of these two components.
As discussed above, the top panel of Fig.~\ref{fig:prof} disentangles
these; and the excess emission revealed is drawn in more detail in
Fig.~\ref{fig:nucprof}.
At radii $\leqslant 0.5$\arcsec\, the data show a clear break at about
0.15\arcsec\ characterised by a change in the radial luminosity gradient that
is independent of the underlying bulge light 
(extrapolated from radii 2--10\arcsec\ and 
marked by a dashed line).
The profile of the excess can be fit by an $r^{1/4}$
profile whose effective radius $r_{\rm eff}$ changes
at 0.10--0.15\arcsec.
However, the two radii are $r_{\rm e1}=0.25$\arcsec\ and 
$r_{\rm e2}=8.2$\arcsec, which are rather larger than the size scales
over which the respective regions extend, leading to some doubt about
whether they are physically meaningful fits.
On the other hand, under the assumption that the excess continuum
arises in a disk, it can be fit equally well by an exponential whose
scale length $r_{\rm d}$ (note 
that disk scale length is related to effective radius by 
$r_{\rm d} = r_{\rm eff} / 1.68$) changes at a radius of 0.11\arcsec.
At smaller radii $r_{\rm d1}=0.037$\arcsec, while at larger
radii $r_{\rm d2}=0.38$\arcsec.

In summary, the radial profile and rotational signature suggest a
disk-like distribution for the nuclear component.
On the other hand, the high dispersion and isophotal position angle
(although both are biassed by the bulge component)
indicate a thicker more spheroidal or even triaxial geometry.
Taken together, these results suggest a thickened disk is the
appropriate interpretation for the nuclear stellar light.

\subsection{Ionised Gas and Young Stars}
\label{sec:brg}

The Br$\gamma$ map in the right hand panel of
Fig.~\ref{fig:fluxmaps} shows only the narrow (FWHM 200--300\kms)
component of the line,
achieved by fitting out the broad ($\sim$3000\kms) component.
The resulting emission is resolved along all PAs.
The crucial question here is whether the bulk of the Br$\gamma$
originates from an outflow in the narrow line region (NLR) or from OB
stars.
On scales of 1--7\arcsec\ the [O\,{\sc iii}] emission is extended on
PAs of $+15^\circ$ to $+30^\circ$ \citep{mun95,sch96}, roughly along
the galaxy's minor axis.
However, \cite{sch96} also note a knot of emission $\sim0.25$\arcsec\ 
from the nucleus that they suggested could be associated with the
double-peaked radio continuum source in \cite{mun95}.
Indeed, approximately 0.2\arcsec\ north and south of the nucleus, the
dispersion of the Br$\gamma$ line is much larger than elsewhere, which
could be indicative that these features are related and originate
in the NLR.
On the other hand, both the morphology and primary velocity gradient
of the Br$\gamma$ are oriented to the north-west -- which is strong
evidence for a direct relation to the stars and molecular gas.
A quantitative discrimination can be made by assigning emission with a
dispersion greater than 150\,kms to the NLR, and the rest (which lies
along the major axis) to star formation.
We find that $\sim75$\% of the Br$\gamma$ flux most probably
originates in star formation.
Section~\ref{sec:stars} and Fig.~\ref{fig:ratios} address the consequences
on our starburst models of the uncertainty of this conclusion.

\subsection{Radio Continuum and Supernova Remnants}
\label{sec:radcont}

A map of the 6\,cm radio continuum at a resolution of
0.05\arcsec$\times$0.07\arcsec, comparable to that we have achieved,
was published by \cite{mun95}.
This showed a 0.3\arcsec\ long structure, composed of knots of
emission, elongated at PA $-10^\circ$.
In the 0.14\arcsec$\times$0.17\arcsec\ resolution 18\,cm map these
features were visible only as a bright compact source and a tail to
the north;  
but an additional bright knot was also apparent 0.4\arcsec\ to the
north.
\cite{mun95} suggested that these features might be radio jets --
either one or, if the nucleus actually sits between the brightest
knots, two sided.
However, they noted a number of difficulties with this interpretation:
lateral extensions were detected well
above the noise level; and the jet orientation differs
from that of the narrow line region, as traced by [O{\sc iii}].
They suggested that as in NGC\,4151 \citep[e.g.][]{ped93}, misalignment
may be expected between the radio axis, which lies along the
collimation axis of the UV ionisation cone, and a density bounded
extended narrow line region (ENLR).
However, they also realised that the line widths of the [O{\sc iii}]
are indicative of a NLR rather than an ENLR, and in addition for such
a model to apply, the NE side of the disk would have to be closer and
hence the spiral arms would have to be leading rather than trailing.
To circumvent these difficulties, they suggested that the radio
collimation and [O{\sc iii}] outflow may not be due to the same
mechanism.

\begin{figure}
\epsscale{0.9}
\plotone{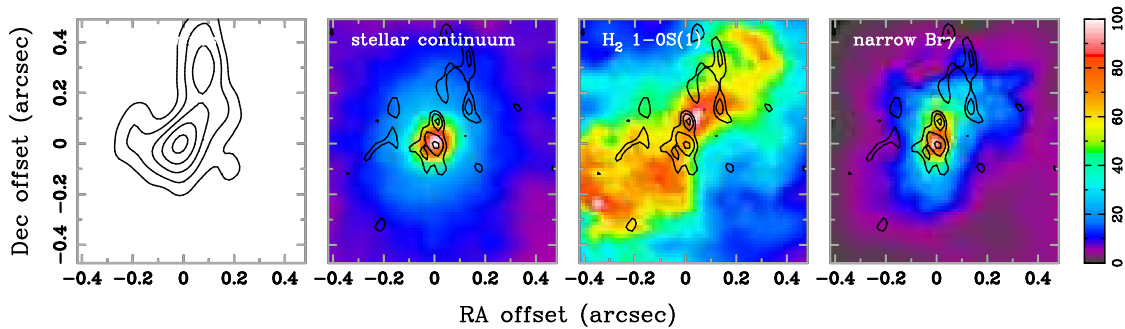}
\caption{Contour plots of the 6\,cm radio continuum kindly provided by
  C.~Mundell.
Left panel: at low resolution (u,v-tapered image with beam size
  $0.191\arcsec\times0.141\arcsec$),
  demonstrating that the 2 blobs seen at 18\,cm \citep{mun95}
  break up into discrete sources at higher resolution.
Right 3 panels: at high resolution (naturally weighted image with beam
  size $0.076\arcsec\times0.053\arcsec$).
Contour levels are at 2, 3.5, and 5 times the rms noise of
  0.37\,mJy\,beam$^{-1}$.
Images on which the contours are superimposed are from
  Fig.~\ref{fig:fluxmaps}.
It is not possible to align the data astrometrically.
Instead, we have assumed that the brightest 6\,cm peak coincides with
  the peak in the K-band stellar continuum -- justifiable under the
  assumption that the radio continuum is due to star formation. This
  alignment is not critical to the interpretation. 
North is up and east is left.
}
\label{fig:radio}
\epsscale{1.0}
\end{figure}

These discrepancies cease to be issues -- indeed, actually are
expected -- if instead one attributes the radio
continuum to supernova remnants (SNRs).
In this scenario, the knotty morphology could arise from the
superposition of many SNRs, some of which are perhaps bright enough to
be detected as individual radio supernovae.
Supporting this, the bright northern knot apparent in the 18\,cm map
is a result of the low resolution of that map:
Fig.~\ref{fig:radio} shows that the same knot seen in a low
resolution 6\,cm map breaks up into discrete structures
at higher resolution.
For such an interpretion, the details of the morphology have no more
significance than tracing the locations of these stochastic events.

The flux density of the individual 6\,cm knots in NGC\,3227 is
variable, with the brightest being 1--2\,mJy (excluding the extended
nuclear knot which probably comprises many SNR).
This implies luminosities up to 3--$6\times10^{19}$\,W\,Hz$^{-1}$, fully
consistent with the range of 1--$1000\times10^{19}$\,W\,Hz$^{-1}$ for
the peak 6\,cm flux density of galactic radio supernovae given by
\cite{wei02}.
For comparison, in Arp\,220, 16 radio supernovae with 18\,cm luminosities
20--$60\times10^{19}$\,W\,Hz$^{-1}$ (a factor $\sim$3 lower at 6\,cm,
assuming a spectral index $\alpha=-1$) were reported by \cite{smi98};
and for the 24 sources monitored in the nucleus of M\,82, 6\,cm
luminosities are in the range 0.1--$10\times10^{19}$\,W\,Hz$^{-1}$
\citep{kro00}.
In both these galaxies, most of the SNRs appear to have varied rather
little in flux over respectively 5 and 12 year 
periods \citep{rov05,kro00}.
The flux densities of the knots in NGC\,3227 are therefore certainly
consistent with their being radio supernovae.

We have superimposed the radio continuum data (naturally weighted,
$0.076\arcsec\times0.053\arcsec$ beam size, kindly made available and
re-reduced by
C.~Mundell) and SINFONI data, as shown in Fig.~\ref{fig:radio}.
Since no astrometrical alignment is possible, we have assumed that the
brightest 6\,cm emission coincides with the peak in the K-band
stellar continuum -- justifiable if, as we suspect, the 6\,cm emission
is due to star formation.
For this interpretation, the exact astrometrical
alignment is not critical.
An alternative alignment could be for the near infrared nucleus to lie
between the two main emission knots seen in the 18\,cm map of
\cite{mun95}.
However, this seems unlikely because, as we have shown, the jet-like
structure is in fact small spread-out knots seen at low resolution. 
There is perhaps some correspondance between these off-nuclear 6\,cm
continuum knots, which tend to lie to the northwest, and the
Br$\gamma$ line which is also more extended in the same direction.
Both of these can be understood in terms of a slight tendency for
more, or more recent, star formation there.
On the other hand, it is clear that there is little one-to-one
correspondance between the 6\,cm continuum and 1-0\,S(1) line. 

To estimate the emission within 0.8\arcsec, we take two extreme
limits: that for the nucleus (i.e. southern component) only, as well
as the total, both given by \cite{mun95}.
We estimate the maximum AGN contribution by measuring the flux in the
central 0.090\arcsec\ (to include the full beam) as no more than 10\%
of the total.
We have made no correction for this.
Using the conversion given in \cite{con92} for these 6\,cm and
18\,cm flux densities, we estimate the supernova rate to lie within
the range 0.008-0.019\,yr$^{-1}$.

\subsection{Star Cluster Models}
\label{sec:stars}

\begin{figure}
\epsscale{0.45}
\plotone{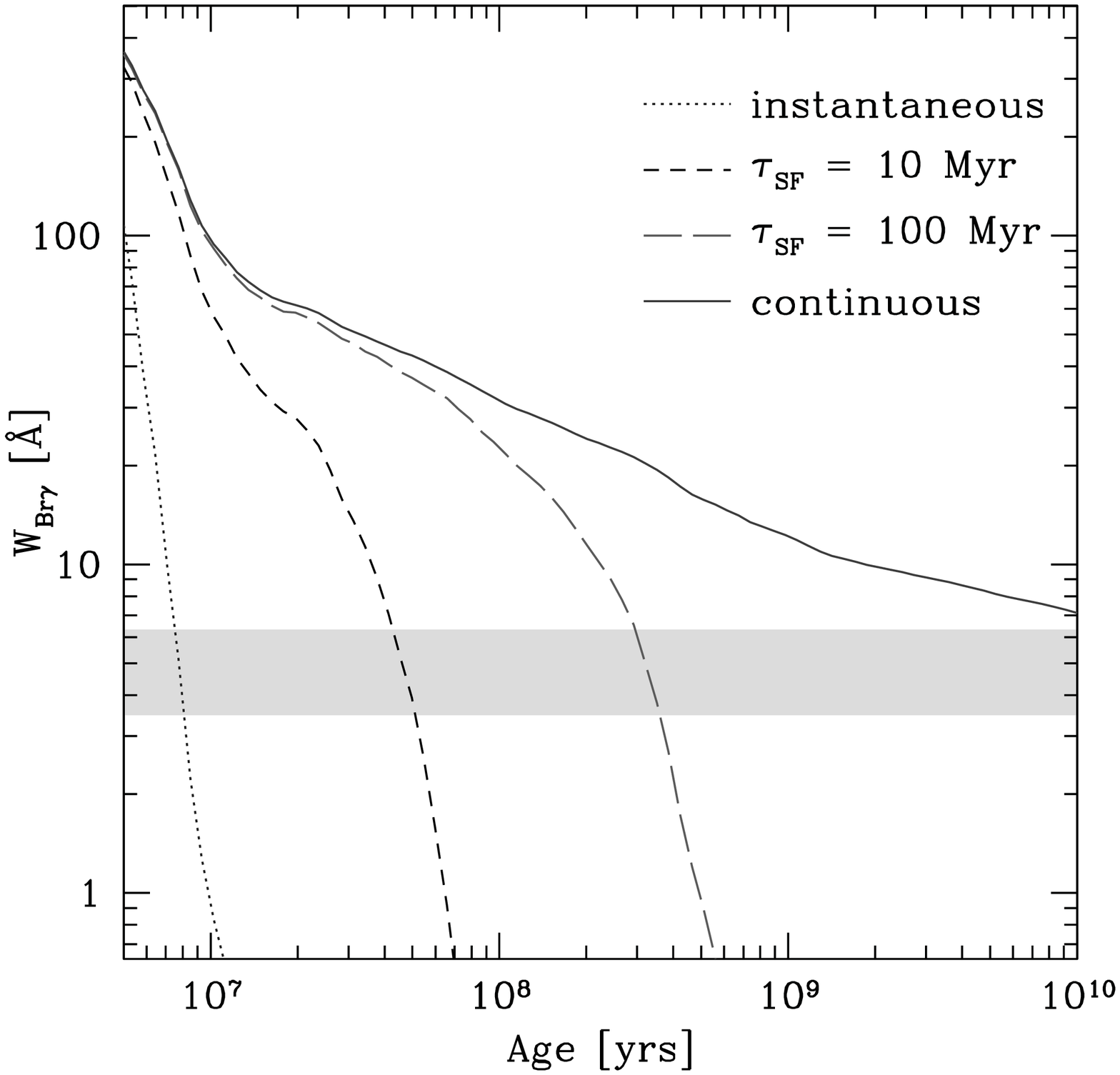}
\plotone{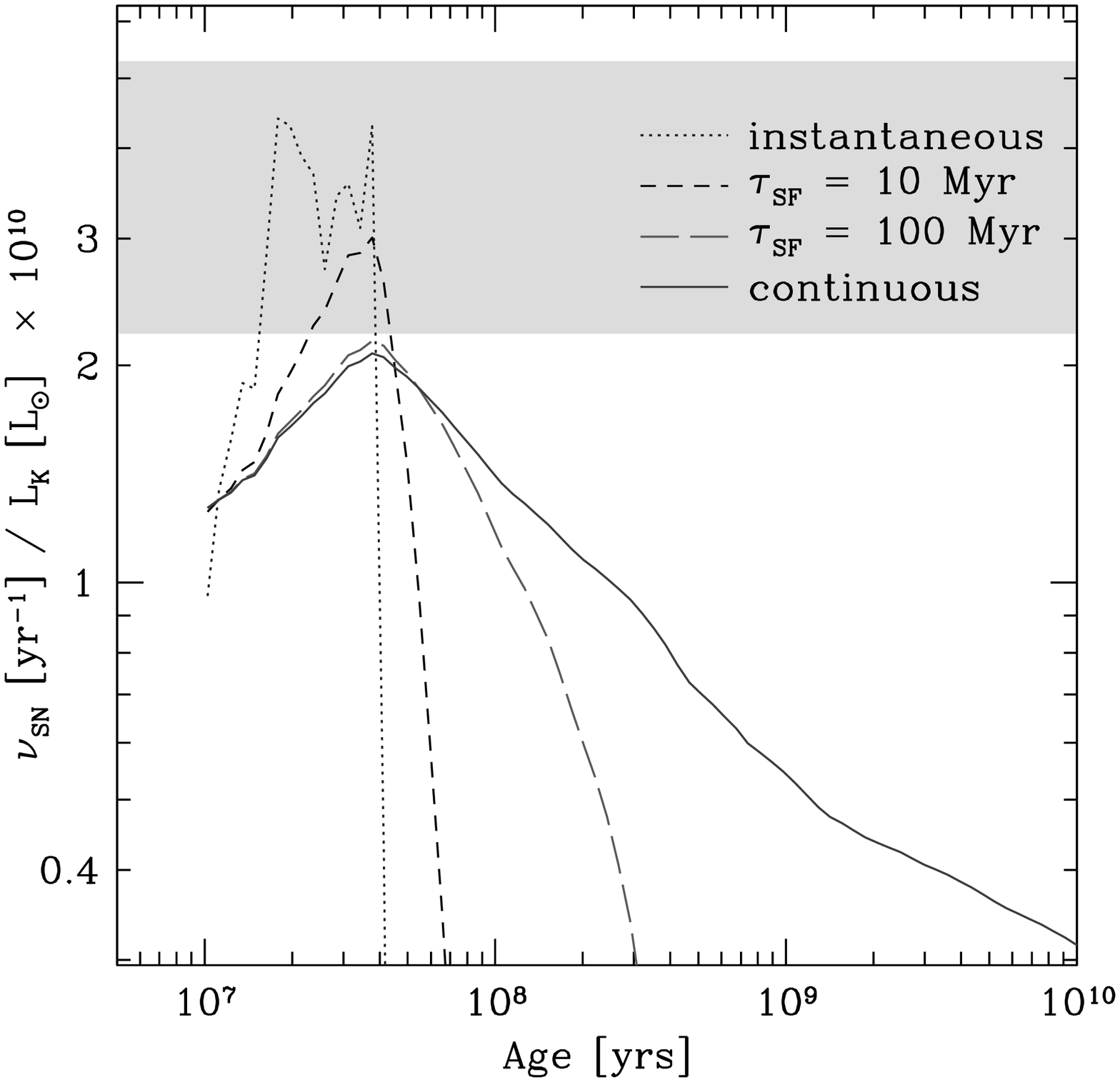}
\caption{Plots of $W_{\rm Br\gamma}$ (left) and $\nu{\rm SN}/L{\rm K}$
  (right) as a function of age for four star formation histories.
The grey bands indicate the range of the measured ratios, and 
in both cases only the stellar continuum has been considered.
The only scenario which yields a consistent age for both diagnostics
  is star formation over a timescale of $\sim$10\,Myr which occured
  $\sim$50\,Myr previously.
}
\label{fig:ratios}
\epsscale{1.0}
\end{figure}

The discussion above has yielded two independent diagnostics which can
be used to constrain the star formation history in the central arcsec
of NGC\,3227.
The first is the (classical) equivalent width of Br$\gamma$.
We use the ratio of the narrow Br$\gamma$ associated with young stars
(i.e. excluding the NLR contribution) to young stellar continuum,
thus removing any bias associated with the AGN or bulge.
The ratio we adopt is therefore $W_{\rm Br\gamma}=5$\,\AA, with an
uncertainty of no more than 1.5\,\AA.
The second is the ratio of the supernova rate to stellar continuum
luminosity, which is 
$10^{10} \, \nu_{\rm SN} (yr^{-1}) \, / \, L_{\rm K} (L_\odot) =2.2$--5.3.
These ratios are presented graphically together with models for various
star formation histories in Fig.~\ref{fig:ratios}, allowing some
conclusions to be drawn immediately.

The $W_{\rm Br\gamma}$ is too low for the star formation to be
continuous, a result supported by the 
$\nu_{\rm SN}/L_{\rm K}$ which is sufficiently large that it permits
only young ages. 
This conclusion is robust even to large uncertainties in the two
parameters.
In particular Fig.~\ref{fig:ratios} shows that the conclusion is valid
even if either $W_{\rm Br\gamma}$ is 
overestimated by insufficient corretion for the NLR contribution, or
if $\nu_{\rm SN}/L_{\rm K}$ is overestimated by not correcting for a
possible AGN component.
Thus any star formation that did occur has now ceased.
On the other hand, because there are already supernova remnants, at
least 10\,Myr must have elapsed since star formation began.
Thus an instantaneous burst is ruled out, because of the steepness
with which $W_{\rm Br\gamma}$ falls.
For the intermediate histories, the age of the star formation is
strongly constrained by $\nu_{\rm SN}/L_{\rm K}$, even if the ratio
itself is rather uncertain, to be at most 50\,Myr.
Even if one allows the uncertainty in the conversion from continuum
flux density to supernova rate, it would be hard to reach ages
greater than 100\,Myr.
With this in mind, the low $W_{\rm Br\gamma}$ requires a short but
finite burst timescale. 
The best model is for a burst with an e-folding decay time of 10\,Myr
which began 40\,Myr ago.
However, we emphasise that a unique match is not the aim of this
modelling;
the important result is that the star formation is very young, and that the
active episode lasted for a finite time but is now finished.
This is in interesting result, begging the questions: since stars form
in cold quiescent environments, how did such intense star formation
occur in this environment which is clearly extremely turbulent (as
evidenced by the 1-0\,S(1) morphology in Fig.~\ref{fig:fluxmaps} and
the velocity dispersion of 100--125\kms, which is significantly higher
than the rotational velocity of 50--100\kms)? 
Once started, what is it that caused the star formation to cease while
the gas is still so plentiful? 
We address these questions in Section~\ref{sec:torus}.

The properties we find for the star formation scenario given above for
the measurements in a 0.8\arcsec\ aperture are
summarised in Table~\ref{tab:derived}.
The star formation has an initial rate of 3\,$M_\odot$\,yr$^{-1}$, 
and a current rate of 0.05\,$M_\odot$\,yr$^{-1}$.
The current mass to K-band light ratio is $M/L_{\rm K} = 0.5
M_\odot/L_\odot$, where the mass of $2.0\times10^7$\,$M_\odot$
refers to the current live stars rather than the total gas consumed
during the active phase which is almost 50\% greater.
Remarkably, the current bolometric luminosity attributable to these
stars is $\log{L_{\rm bol}/L_\odot} = 9.5$. 
This is a crucial result, since it implies that the extinction to the
stars cannot be very great.

Fig~\ref{fig:nucprof} shows that within the central 0.8\arcsec\ we
discuss above,
the intensity increases dramatically at $r<0.11$\arcsec\ (9\,pc).
Within this smaller radius, the star formation models show that the mean
stellar mass surface density is $3\times10^4 M_\odot$\,pc$^{-2}$.
Averaged over the last 40\,Myr (i.e. since the burst began) the mean
absolute star formation rate is only 0.13\,$M_\odot$\,yr$^{-1}$, but the
rate per unit area has been phenomenally high:
typically 500\,$M_\odot$\,yr$^{-1}$\,kpc$^{-2}$, and reaching rates 10
times higher intensity at its peak when it was active
This is more like what one typically expects of ultraluminous
galaxies, and suggests 
that during active star forming phases, which appear to last for
similar timescales of a few 10$^7$\,yr, the local
(i.e. $\lesssim$10--100\,pc) environment around AGN is comparable to
that in ULIRGs.

\subsection{Extinction}
\label{sec:extinc}

So far we have implicitly assumed that there is negligible
extinction, and already we account for 20\% of the bolometric
(i.e. 0.3--1000\micron) luminosity of the entire galaxy.

If we take the reasonable stance that the gas and stars are mixed
(rather than the gas lying in front of the stars as a foreground
screen),
then when considering the effects of extinction we should use the
appropriate mixed model for which the reduction in observed intensity
depends on optical depth through the gas and stars as 
$I/I_0 = (1-e^{-\tau_{\rm mix}})/\tau_{\rm mix}$.
The usual scaling of the optical depth with wavelength 
$\tau_{\rm K} = 0.1 \tau_{\rm V}$ still applies, as does the
definition of the reddening
$A_\lambda = -2.5\log{I_\lambda/I_{\lambda0}}$.
This means that in contrast to the screen model for which 
$A_\lambda = 1.09\tau_\lambda$, in the mixed model a modest observable
reddening can hide the existence of a very large optical depth of gas
and dust since
$A_\lambda = 2.5\log{\tau_\lambda}$ (for $\tau_\lambda \gtrsim 3$).
Note, however, that often extinctions for mixed models are given as
a sort of `screen equivalent', giving rise to very large values.
To avoid this confusion when discussing flux attenutation, in the
remainder of this paper we refer to the more physical
quantity optical depth $\tau_\lambda$ rather than the observationally
motivated extinction $A_\lambda$.

As an illustration, a reasonable optical depth of
$\tau_{\rm V} = 10$ for the mixed model would
mean that 90\% of the UV and optical light, which dominate the spectral
energy distribution, would be re-radiated in the mid- to far-infrared.
Additionally, the scale of the starburst would increase since
$L_{\rm K}$ becomes 1.6 times that observed.
Thus the net effect is that the bolometric luminosity of the starburst would 
be $L_{\rm bol}/L_\odot = 9.7$ and account for approximately 50\% of
the infrared (8--1000\micron) luminosity (although none of the optical
luminosity).
Similarly, one can derive the maximum possible mixed model optical depth,
for which the starburst accounts for all the 8--1000\micron\
luminosity, to be $\tau_{\rm V} = 26$.
This is a strong constraint on the extinction, and we return to this
point in Section~\ref{sec:torusgas}.

\begin{figure}
\epsscale{0.8}
\plotone{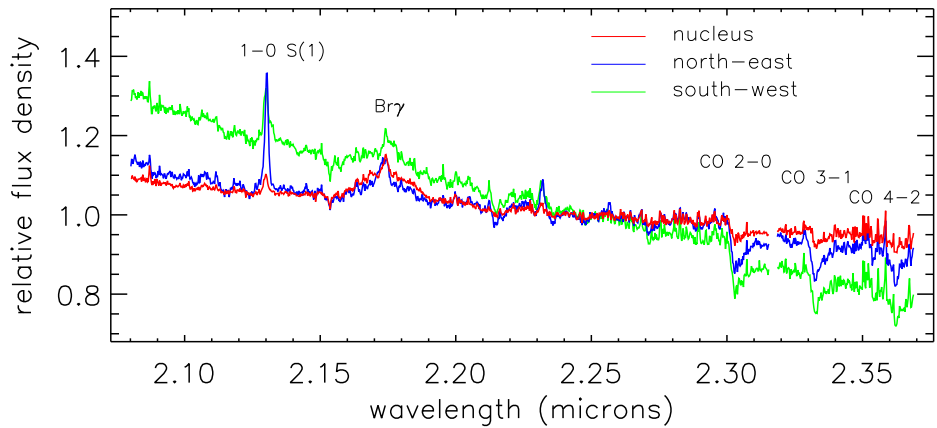}
\caption{Spectra extracted from the SINFONI data cube in 0.25\arcsec\
  apertures centered on the nucleus (red), 0.25\arcsec\ to the
  north-east (blue), and 0.25\arcsec\ to the south-west (green).
All spectra are normalised at 2.25\micron.
The effect of dilution on the CO bandheads is clearly seen in the
  nuclear spectrum.
}
\label{fig:spec}
\epsscale{1.0}
\end{figure}

Do we see evidence for any extinction?
The F160W image in Fig.~\ref{fig:hst_div} highlights a curious feature
in the central arcsec.
In addition to the bright point source associated with the AGN, there
is a strong contrast in brightness between the SW and NE sides of the
nucleus -- corresponding respectively to near and far sides of the
galactic disk for the orientation above.
This is reflected in the corresponding spectra extracted from the
SINFONI data cube, as Fig.~\ref{fig:spec} shows.
The nuclear spectrum is characterised by a flat slope and shallow CO
bandheads indicative of dilution by hot dust and perhaps also extinction.
The spectrum to the south-west has a bluer slope and deeper bandheads,
both consistent with pure un-reddened stellar light.
In contrast, the north-eastern spectrum has deep bandheads but a
flatter slope, suggesting it is reddened (but not diluted).
A differential extinction (screen model) of $A_{\rm K}=1.2$ would
produce the observed change in spectral slope.
On the other hand, for the mixed model, any (largish) extinction is
possible.
For a mixed model, the differential extinction between 2.1 and
2.3\micron\ saturates at this level, and cannot make the slope redder.
One might conclude that while the star formation on the south west
side might be predominantly unobscured, that on the north east side is
probably mixed with considerable dust and gas.
Conversely on larger scales of 1--2\arcsec, \cite{cha00} found
evidence from optical and infrared colour maps for greater extinction
on the south west side.


\section{Stellar Kinematics and the Black Hole Mass}
\label{sec:st_kin}

\begin{figure}
\epsscale{0.6}
\plotone{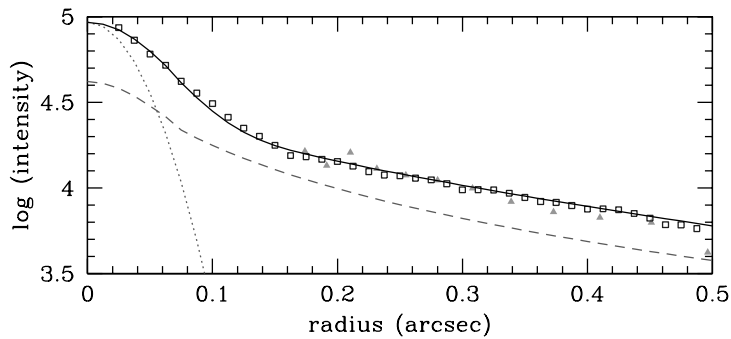}
\caption{Radial profile of the nuclear stellar light.
Filled triangles denote HST F160W data, and open squares the SINFONI
K-band stellar continuum.
The grey dotted line indicates the spatial resolution by tracing a
Guassian with 0.085\arcsec\ FWHM.
The grey dashed line represents the $r^{1/4}$ bulge fit at
2--10\arcsec\ radii extrapolated inwards.
The dark solid line indicates a fit to the data including the bulge
component and an exponential whose scale length changes at a radius of
0.11\arcsec.
}
\label{fig:nucprof}
\epsscale{1.0}
\end{figure}

\begin{figure}
\epsscale{0.4}
\plotone{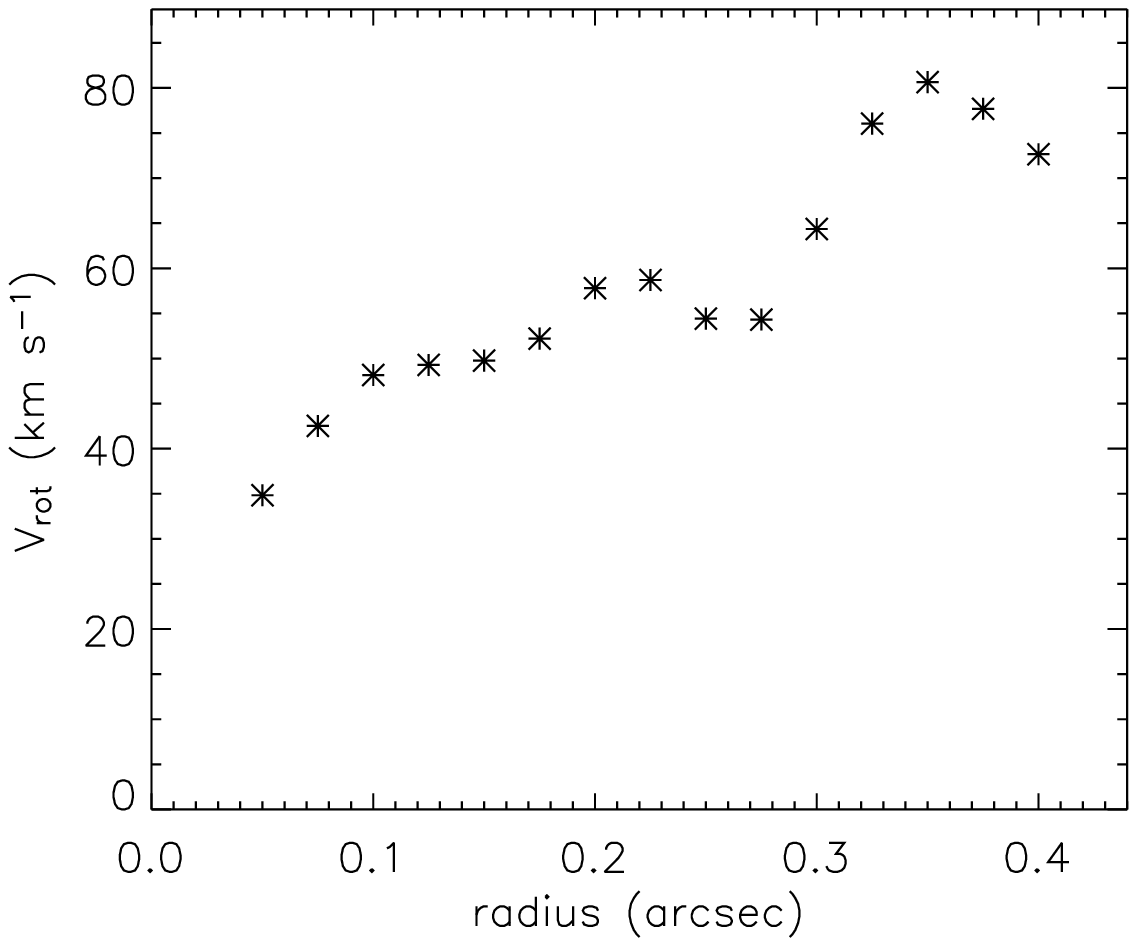}
\plotone{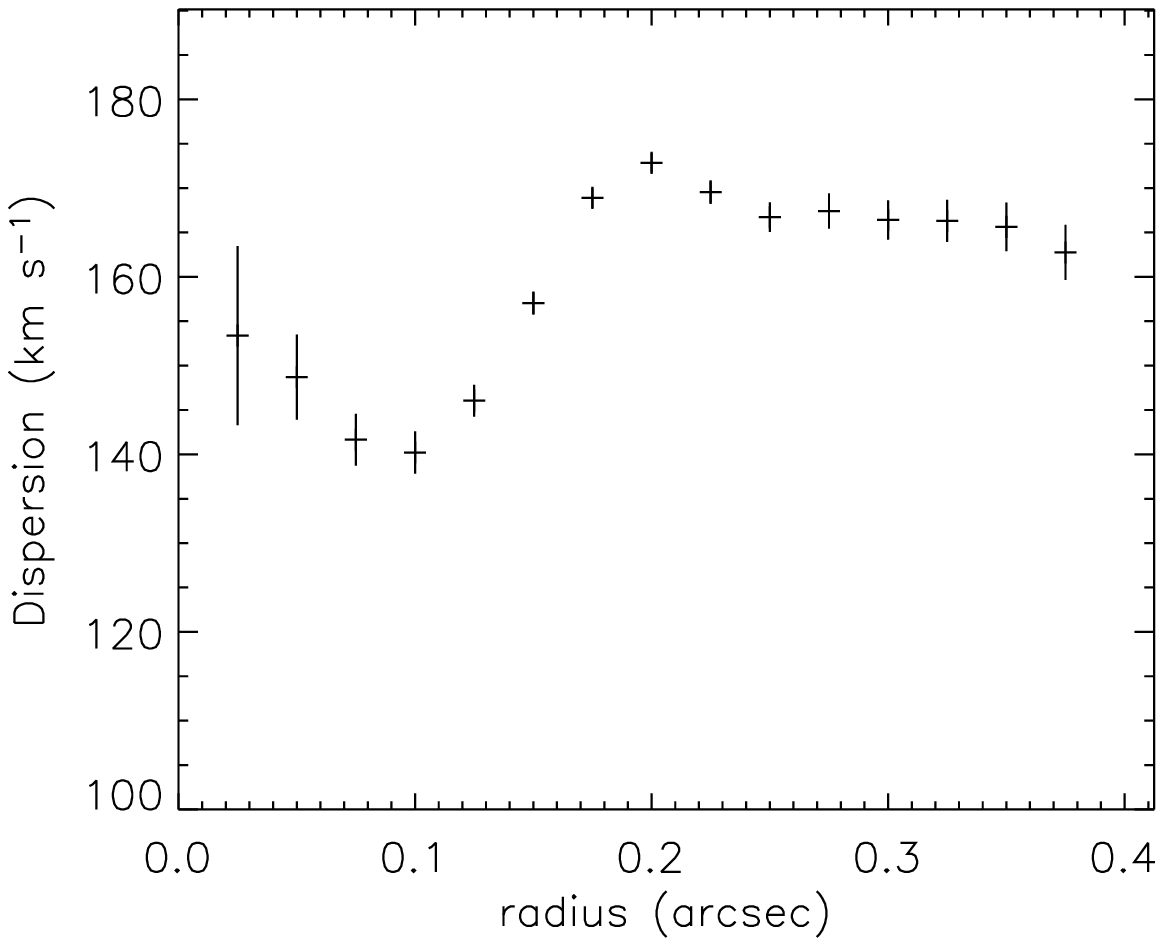}
\caption{Left: stellar rotation curve (found using the commonly employed
  tilted ring model put forward by \citealt{beg89}); Right: 
azimuthally averaged radial profile of the stellar dispersion.
In both cases a fixed inclination of 55$^\circ$ at position angle
135$^\circ$ was adopted.
The velocity increases steadily to $\sim$80\kms\ at 0.4\arcsec.
The dispersion of 160--170\kms\ drops quickly to 140\kms\ inside a radius of
0.2\arcsec\ -- perhaps because of the increasing predominance of the
young stars; and then begins to increases slowly at the smallest
radii -- perhaps due to the dynamical impact of the black hole.
}
\label{fig:vrot_disp}
\epsscale{1.0}
\end{figure}

None of the velocity fields in Fig.~\ref{fig:velfield} show
evidence for the extreme warp proposed by 
\cite{sch00} based on their 0.6\arcsec\ resolution CO\,(2-1) data.
The existence of a small warp would not be surprising since
very high spatial resolution 
mapping of masers indicate that on small spatial scales warps may be
common in galaxy nuclei -- for example 
NGC\,4258 \citep{her96}, 
NGC\,1068 \citep{gre96}, and
Circinus \citep{gre03}.
The reason for the apparent discrepancy is that \citealt{sch00}
had interpreted the high gas dispersion as spatially
unresolved rotation or inflow -- whereas our data
show that the gas dispersion $\sigma_{\rm gas} \sim 120$\kms\ is
intrinsically high.
This is not very different from the stellar dispersion $\sigma_*$ of
140--160\kms, which indicates that random motions rather than ordered
rotation dominate the kinematics of the stars.
As indicated above, the observed data are the superposition of
kinematics of the bulge and nuclear component, and this may be
responsible for some of the unusual 
characteristics of the dispersion most easily seen in
Fig.~\ref{fig:vrot_disp}.
This shows that $\sigma_*$ is constant at larger radii (out to our
radial limit) but decreases at 0.1--0.2\arcsec.
This is exactly the radius at which Fig.~\ref{fig:nucprof} shows an
increase in the intensity of the stellar light from the nuclear
component.
It is therefore possible that at larger radii the observable $\sigma_*$
is dominated by the older bulge, but at smaller radii $\sigma_*$ is more
strongly affected by the stars in the nuclear region.
If this is indeed the case, then the subsequent increase of
$\sigma_*$ very close to the nucleus could be due to the influence of
the black hole.

To test whether the behaviour of $\sigma_*$ is consistent with what
one might expect from the nuclear star cluster and black hole, we have
made a simple dynamical model.
This is realised as an edge-on thin disk for which the mass profile
follows the luminosity profile
given in Fig.~\ref{fig:nucprof}, and accounts for both the gas and
stellar mass.
For this purpose, we have ignored the luminosity from the large scale
bulge (denoted by the dashed line in the figure).
To simulate the random motions which we have shown are important, we
have convolved the model spectrally with a Gaussian having
$\sigma=120$\kms.
The model is then convolved with the effective spatial resolution and
finally `observed' (see \citealt{dav04a,dav04b} for more
details of how the model is generated).
Although this model does not represent the true 3-dimensional mass
distribution, the radial distribution is correct and hence it is able
to provide a first estimate of the way in which the velocity
dispersion changes with radius.
In addition it can, at least quantitatively, show how the rotation
velocity is expected to change with radius.
The first conclusion is that the increase in $\sigma_*$ at
$r<0.1$\arcsec\ is indeed consistent with the nuclear mass
distribution and presence of a black hole with mass of order
$2\times10^7$\,M$_\odot$.
The second conclusion confirms that, as expected, the system cannot
be rotationally supported: the measured velocities in
Fig.~\ref{fig:vrot_disp} are only 60--65\% of those needed for this.
Nevertheless, the steep increase in V$_{\rm rot}$ out to
$r=0.1$\arcsec\ and subsequent more gradual rise do match the shape of
the model rotation curve.

In order to understand the kinematic behaviour more fully and to
determine $M_{\rm BH}$ from the stellar kinematics, we have
constructed an appropriate Schwarzschild orbit superposition model,
which is described and discussed below.
NGC\,3227 is a good candidate for such an analysis because it has already
been the focus of other efforts to derive $M_{\rm BH}$.
These include reverberation mapping, which yielded
$4.2\pm2.1\times10^7$\,M$_\odot$ \citep{pet04};
X-ray variability measurements, giving 
$2.2\times10^7$\,M$_\odot$ \citep{nik04};
and the $M_{\rm BH}-\sigma_*$ relation itself, leading to 
$3.6\pm1.4\times10^7$\,M$_\odot$ \citep{nel04}.
Based on these masses and the velocity dispersion of 136\,\kms\
measured by \citep{nel04}, one can estimate the `radius of influence'
of the black hole to be 
$r_{\rm g} \sim G\,M_{\rm BH}/\sigma^2 \sim $8\,pc.
Hence the black hole dominates the dynamics in the central 16\,pc, a
region which is easily resolved at our 7\,pc FWHM spatial resolution.

%
%
\subsection{Schwarzschild Modelling}
\label{sec:schwarz} 

\citet{sch79} orbit superposition technique has become
the standard tool for deriving M$_{\rm BH}$ from the kinematics
of the surrounding stellar spheroid. The procedure commonly 
involves four steps: (1) the photometry is deprojected to
get the luminosity distribution $\nu$ of the stars; (2) a gravitational
potential is constructed from trial values for the stellar mass-to-light ratio
$\Upsilon$ and the black hole mass $M_\mathrm{BH}$; (3) 
thousands of orbits in this potential are
combined to match the stellar luminosity distribution and the kinematical 
constraints; (4) $\Upsilon$ and $M_\mathrm{BH}$ are systematically varied to 
find the optimal solution in a $\chi^2$ sense.
The specific implementation of the method
we use here is detailed in \cite{tho04}. The program is based on the code of the Nuker 
team (Richstone et al. in preparation) that has been used
to measure black hole masses in numerous elliptical galaxies
(\citealt{geb03}), including Cen\,A \citep{sil05}. 
We apply these models here with the aim of constraining the black hole
mass rather than necessarily deriving uniquely the kinematic structure
of the entire system.

\begin{figure}
\epsscale{0.8}
\plotone{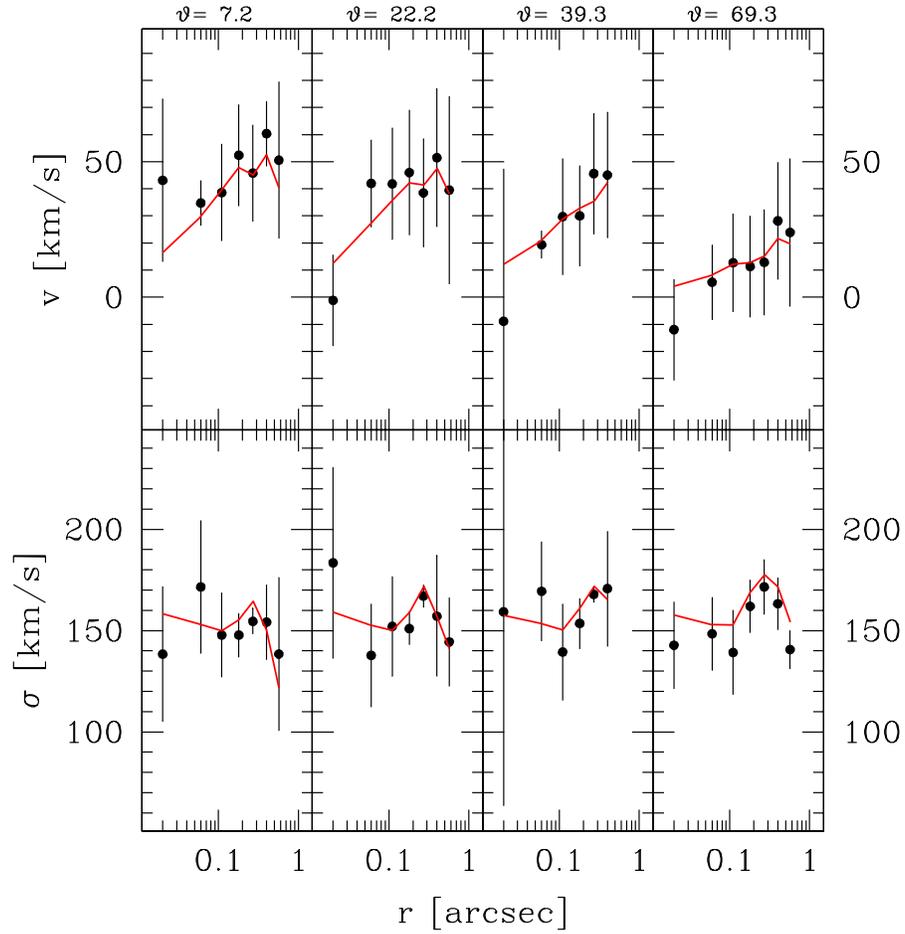}
\caption{Comparison of model and data velocity (upper) and dispersion (lower). 
The 4 panels from left to right represent the 4 angular bins (from
  major axis at left to minor axis at right), the
  mid-points of which are indicated at top.}
\label{fig:ghplot}
\epsscale{1.0}
\end{figure}

The validity of our models rests on two major assumptions: that the
central region of NGC\,3227 is (1) stationary and (2) axisymmetric. 
The first assumption is difficult to verify independently. From the quality
of our fits to the data (cf. Fig.~\ref{fig:ghplot}) we can however conclude
that -- within the uncertainties -- this assumption is consistent with
the data. 
To which degree the second assumption holds for NGC\,3227 can be
estimated from variations in the 
kinematics from quadrant to quadrant (cf. Sec.~\ref{input:kin}). 
Providing means to measure
such variations is one of the advantages of using a full 2-dimensional
field of view as we do here. 
The other advantage is that such data constrain the kinematics of 
all kinds of orbits that are needed to map the system.

\begin{figure}
\epsscale{0.5}
\plotone{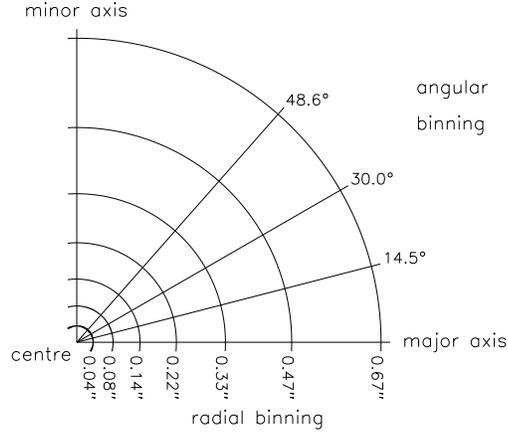}
\caption{Radial and angular binning scheme used as input for the
  Schwarzschild orbit superposition modelling. The same arrangement
  was repeated for each of the four quadrants. The total number of
  bins ($32\times4$) is well matched to the number of indepedent data
  points, given the spatial resolution of 0.085\arcsec\ and the
  0.93\arcsec\ field of view.
}
\label{fig:bins}
\epsscale{1.0}
\end{figure}

The present models for NGC\,3227 are calculated on a grid with 7 radial
and 4 angular bins in each quadrant (Fig.~\ref{fig:bins}), 
giving a total which is comparable to 
the number of spatially independent regions in the data. The 
binning resolution is matched to sample more frequently the 
regions where the greatest variations in $V$ and $\sigma$ are 
expected to lie. For each mass model  about $2\times3300$
orbits are tracked (the initial factor 2 due to the inclusion
of a prograde and retrograde version of each orbit) which, based on
the criterion in \cite{ric04} is easily sufficient.

We now discuss how the observations are prepared for the models.

\subsubsection{Photometry}
\label{input:photo}

As we have shown, the luminosity distribution comprises two parts: the
bulge and the nuclear star forming region.

The bulge, which has an $r^{1/4}$ profile with an effective radius in
the K-band of 3.4\arcsec, contributes 
40\% of the nuclear K-band luminosity and outside of this immediate
region dominates the K-band luminosity to a radius of 9\arcsec.
Due to its high mass-to-light ratio ($M/L_K$), it has an important
effect on the model.
We are unable to constrain its $M/L_K$ ratio directly, but from a literature
search \cite{for03} found that empirical determinations lie in the
range 10--30\,M$_\odot$/L$_\odot$ -- similar to those predicted by
population synthesis models for old ($\sim10$\,Gyr) populations.

The nuclear star forming region contributes 
60\% of the nuclear K-band luminosity and has a
K-band luminosity profile that 
is well matched by an exponential profile with a change in scale
length from 0.037\arcsec\ to 0.38\arcsec\ at 0.11\arcsec.
Its $M/L_K$ ratio is well constrained via population synthesis models
to be 0.5\,M$_\odot$/L$_\odot$.
However, its mass could be dwarfed by that of the molecular gas which
lies in the same region and which we assume has the same radial
profile.
The gas clearly contributes significantly to the gravitational
potential, and implies that the effective $M/L_K$ ratio
should probably be rather higher.

As a novel feature of our models, and due to the uncertainties in the
$M/L_K$ ratios, we allow the contributions
of the two components to the total (stellar) mass-profile $\rho_\mathrm{star}$ 
to be varied independently. 
To this end the nucleus and the bulge
are deprojected separately using the program of \cite{mag99} and
combined via
\begin{equation}
\rho_\mathrm{star} \equiv \Upsilon_\mathrm{nuc} \, \nu_\mathrm{nuc} + 
\Upsilon_\mathrm{bul} \, \nu_\mathrm{bul}.
\end{equation}
Thereby, the (K-band) mass-to-light ratios
$\Upsilon_\mathrm{nuc}$ and $\Upsilon_\mathrm{bul}$ of nucleus and bulge
($\nu_\mathrm{nuc}$ and $\nu_\mathrm{bul}$ are the corresponding
deprojections) are assumed constant with radius. 
We set the ellipticity of nucleus and bulge 
$\epsilon_\mathrm{nuc} \equiv \epsilon_\mathrm{bul} = 0.3$. The
inclination can be varied arbitrarily, but
must be equal for both components (to ensure axisymmetry).

\subsubsection{Kinematics}
\label{input:kin}

The velocity and dispersion were found for each
radial and angular bin shown in Fig.~\ref{fig:bins}, by calculating
the mean of all spatial pixels within each bin, weighted according to
the uncertainty of each measurement.
The uncertainties were combined in a similar fashion to estimate the
standard error of these means. 
The position angle which defines the major axis, and hence the
orientation of the bins, is relatively well constrained.
We have shown that on scales of 140\,pc and more,
there is no doubt that it is at $-30^\circ$.
On the scales we consider here, less than 50\,pc, the data indicate a
preference towards $-45^\circ$.
We have therefore adopted $-40^\circ$, which is sufficiently close to
both limits that, given the size of the angular bins, the small
uncertainty will have no impact on the resulting model.
As input to the orbit models binned line profiles are
generated from the measured kinematical parameters (see, for example, \citealt{tho05}). 
Uncertainties in these parameters are propagated on the basis of Monte Carlo simulations.

The stellar dynamical determination of the black hole mass for NGC\,3227 suffers from two
difficulties: Firstly, due to the constraints of signal-to-noise, it was not possible to
derive the Hermite terms $h_3$ and $h_4$ from the spectra in any
meaningful way. These higher order moments, however, contain
important information about the distribution of stellar orbits \citep{deh93}.
Unfortunately, due to the so-called ``mass-anisotropy degeneracy'' 
uncertainties in the orbit distribution
directly translate into uncertainties in the derived black-hole masses. Anyway,
in order not to bias our models we allow for a rather large range
of $h_3$ and $h_4$ by using $h_3 = h_4 = 0 \pm 0.1$ when
deriving the line profiles and their uncertainties.

The second limitation originates in
the quadrant-to-quadrant variations of $v$ and $\sigma$, which are
generally larger than the statistical errors. This 
indicates that the modelled region is not exactly axisymmetric. When averaging the kinematics
from the four quadrants these variations are taken into account in the error bars.

In summary, the assigned error bars to the kinematical input parameters are rather large
and mostly reflect (1) systematic uncertainties related to central deviations from axisymmetry
and (2) our ignorance about the higher order Gauss-Hermite moments.

\subsection{The Black Hole Mass of NGC\,3227}
\label{results:orbits}

\begin{figure}
\epsscale{0.6}
\plotone{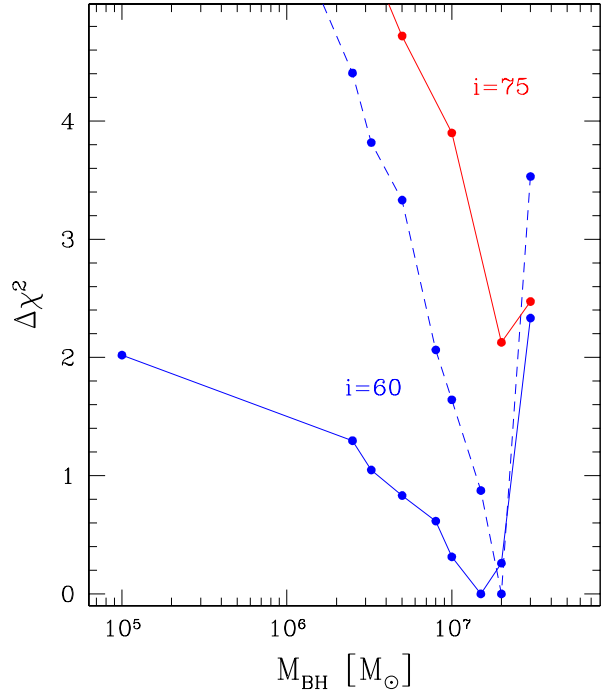}
\caption{Goodness of fit $\Delta \chi^2=\chi^2-\mathrm{min}(\chi^2)$ 
versus black-hole mass (allowing the mass-to-light ratios of both the
nuclear and bulge components to be optimised in the model each time,
irrespective of external constraints):
$i=75^\circ$ (red solid) and
$i=60^\circ$ (blue solid). The blue dashed line shows the case for
rescaled error bars in $h_3$ and $h_4$ (see text for details).}
\label{fig:chibh}
\epsscale{1.0}
\end{figure}

\begin{figure}
\epsscale{0.32}
\plotone{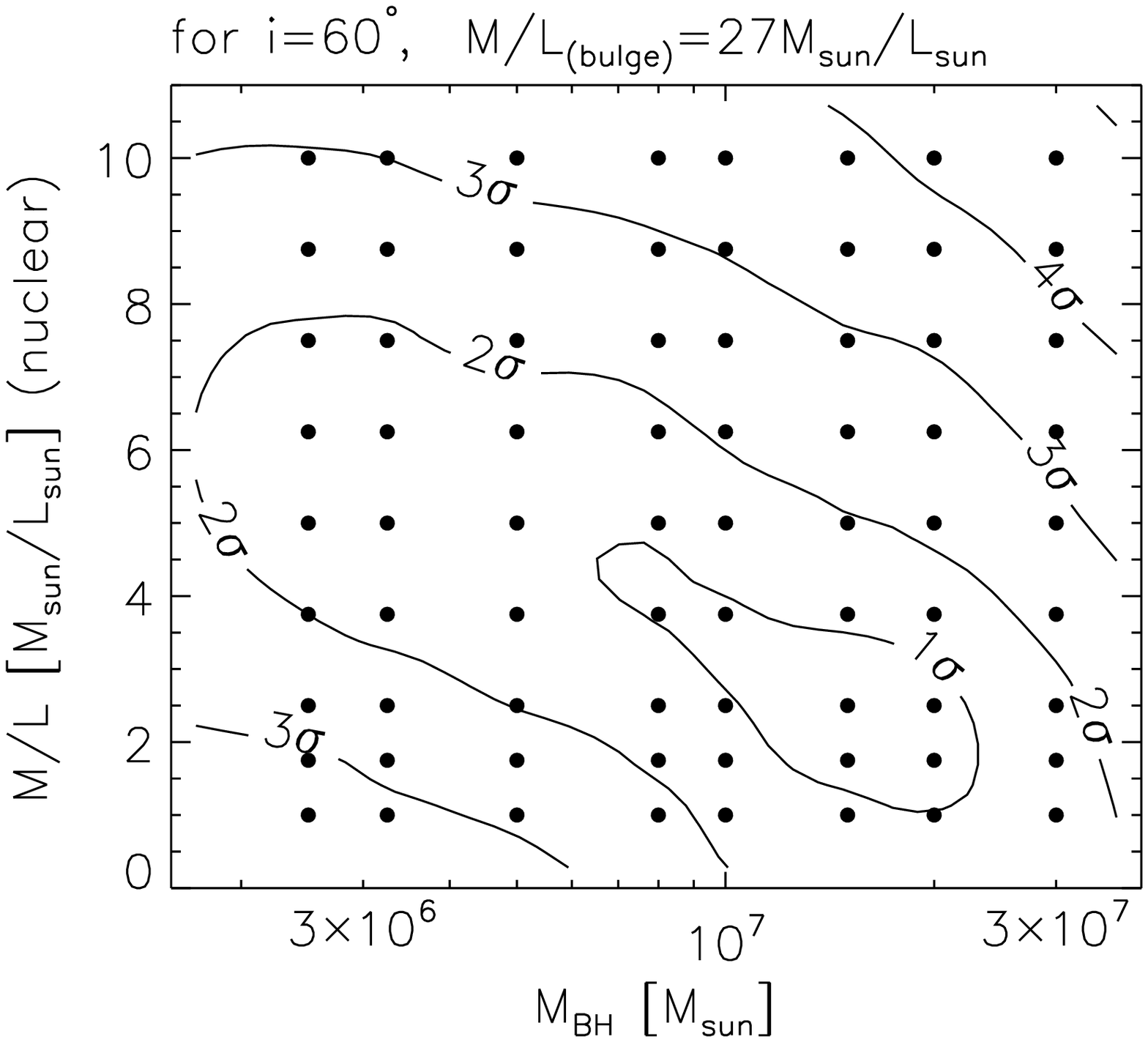}
\plotone{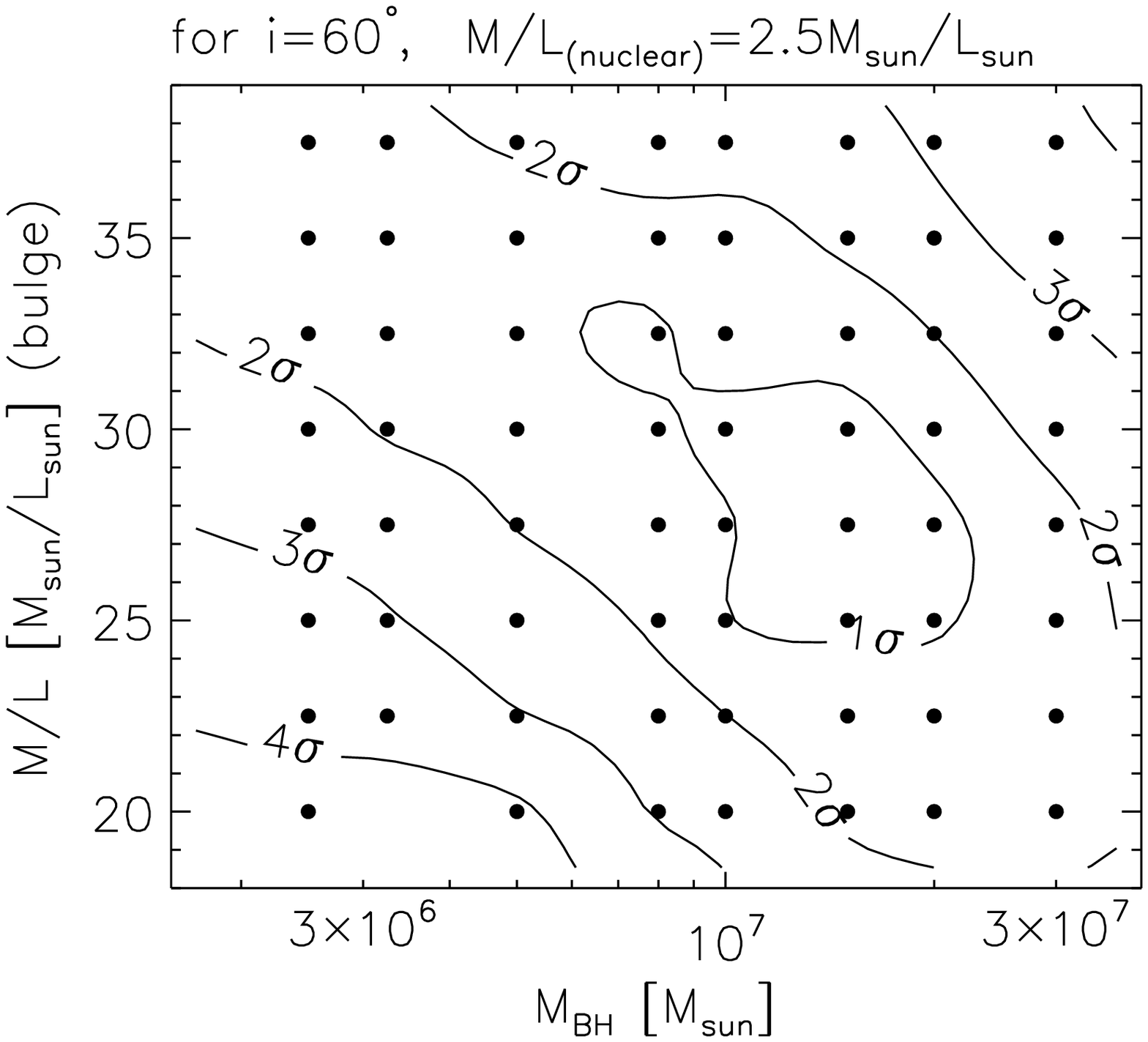}
\plotone{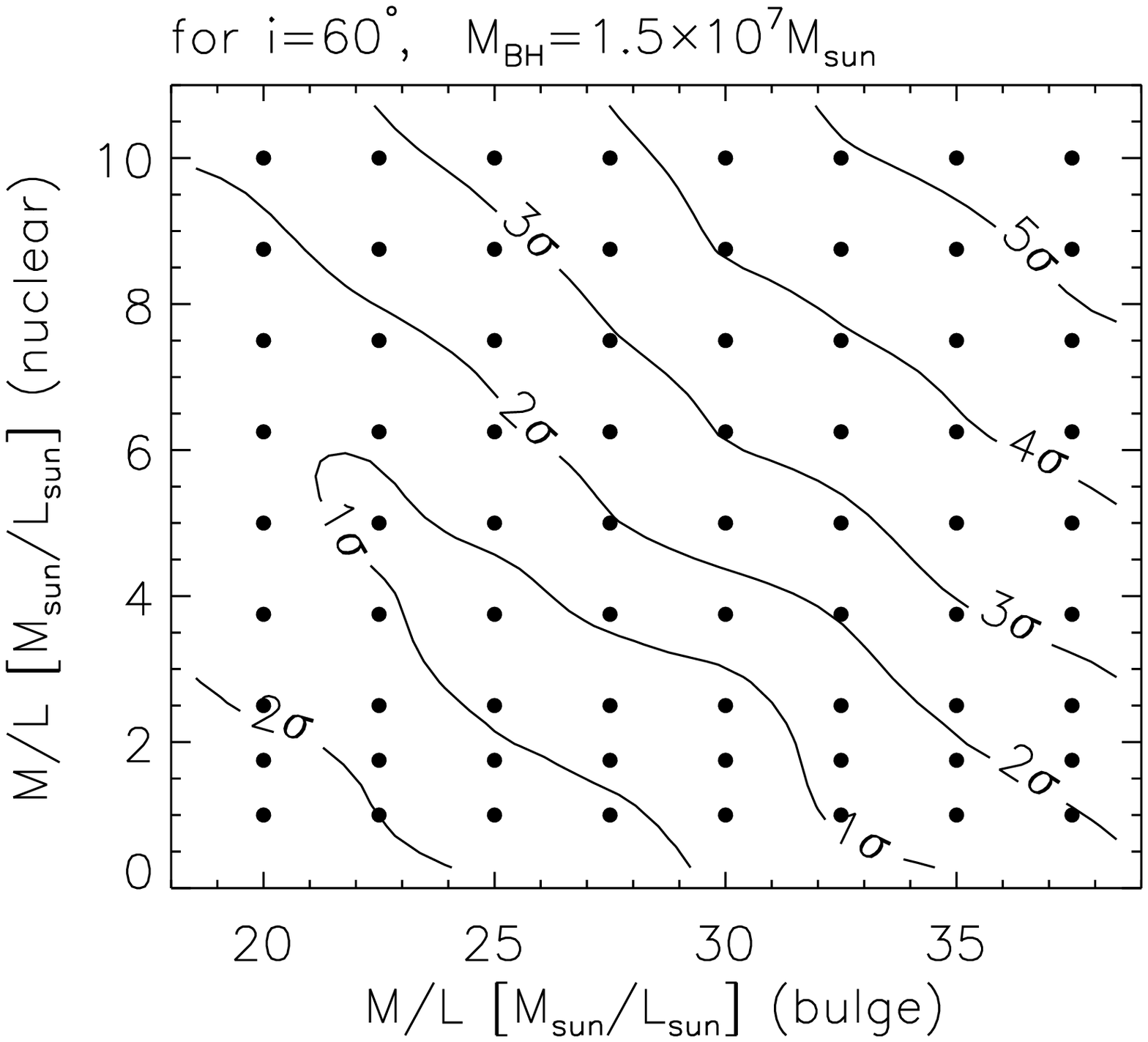}
\caption{The range of models calculated (black points) and inferred
  confidence intervals (contours, in the range 1--5$\sigma$
  corresponding to $\Delta\chi^2=\{1,4,9,16,25\}$) for the
  Schwarzschild models calculated for the best fitting inclination of
  $60^\circ$, with the third parameter in each case fixed at its
  optimal value.
Left: as a function of M$_{\rm BH}$ and Mass/Light ratio of nuclear
  component;
Centre: as a function of M$_{\rm BH}$ and Mass/Light ratio of bulge.
Right: as a function of Mass/Light ratios of the bulge and nuclear
  component.
}
\label{fig:chi2}
\epsscale{1.0}
\end{figure}

\begin{figure}
\epsscale{0.6}
\plotone{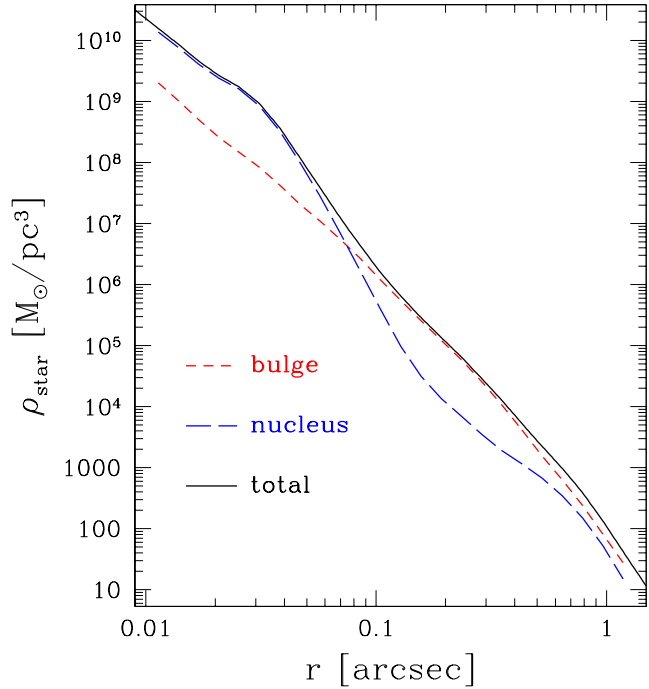}
\caption{Best-fit stellar mass density and its decomposition into the
bulge and nuclear component, respectively.}
\label{fig:rhostar}
\epsscale{1.0}
\end{figure}

The results of the modelling are shown in
Figs.~\ref{fig:chibh}--\ref{fig:ghplot}.
The best fitting model has an inclination of $i=60^\circ$ (fully consistent
with that of the isophotes) and its 
black hole mass is M$_{\rm BH}=1.5\times10^7$ (Fig.~\ref{fig:chibh}). 
The corresponding stellar mass-to-light ratios are $\Upsilon_\mathrm{bul}=27.5$
and $\Upsilon_\mathrm{nuc}=2.5$, respectively. As Fig.~\ref{fig:chibh}
shows, the difference between $i=60^\circ$ and $i=75^\circ$ is formally only
at the one sigma level. In any case, the best-fit black-hole mass does
not depend strongly on the assumed inclination.

It turned out after the modeling that the uncertainties assigned to $h_3$ and $h_4$
probably overestimate the actual freedom in the orbit distribution. From all calculated
models we found the 68 percentiles $\chi^2_{h3}/N_\mathrm{data} = 0.01$
and $\chi^2_{h4}/N_\mathrm{data} = 0.06$, respectively. Consequently, the variations
in the fitted models are much smaller than the orignially allowed 
$\Delta h_3=\Delta h_4 \equiv 0.1$. To illustrate the effect this has on the derived
confidence intervals we have rescaled these
error bars to a third of their original value and recalculated $\Delta \chi^2$. The result
for the case $i=60^\circ$ is shown as the dashed line in Fig.~\ref{fig:chibh}.
As apparent, it does not alter the best-fit black-hole mass
significantly, but changes the confidence intervals drastically.

In the discussion above, we have allowed the Schwarzschild model to
find the best mass to light ratios as free parameters.
On the other hand, as previously intimated, these are in fact already
limited by external constraints based on other arguments,
leaving M$_{\rm BH}$ as the only truly free parameter.
In particular $M/L_K$ for the bulge is unlikely to exceed
30\,M$_\odot$/L$_\odot$ \citep{for03};
and the effective (stars plus gas) $M/L_K$ for the nuclear component
is likely to be in the range 1--5M$_\odot$/L$_\odot$
(Sec.~\ref{sec:torusgas}).
It is for this reason that in Fig.~\ref{fig:chi2}, which shows the
dependencies between the three dynamical parameters, we have calculated
confidence intervals for each pair of parameters assuming that the
third is fixed.
The combination of the degeneracies and the external constraints puts
stronger limits on the range of possible black hole masses than the
Schwarzschild modelling alone.

The figure shows a degeneracy between $\Upsilon_\mathrm{nuc}$ and
M$_{\rm BH}$. Black hole masses of 
M$_{\rm BH}=7 \times 10^6-2\times 10^7$\,M$_\odot$ can be fit with 
$\Upsilon_\mathrm{nuc} = 1-5 \, M_\odot/L_\odot$. 
The reason for the degeneracy is illustrated in
Fig.~\ref{fig:rhostar}: the nuclear component dominates the stellar
mass density in the centre inside 0.1\arcsec\ and a larger
$\Upsilon_\mathrm{nuc}$ can therefore compensate  
a lower M$_{\rm BH}$. 
Since its contribution to the central mass density is lower,
the bulge $\Upsilon_\mathrm{bul}$ is less dependent on the black-hole mass.
But due to their similar light-profiles around 1.0\arcsec, the bulge
and nuclear mass-to-light ratios are coupled strongly to each other.
From Fig.~\ref{fig:chi2}, we find that the $M/L_K$ ratio for the bulge
should lie 
in the range $\Upsilon_\mathrm{bul} = 25-35 \, M_\odot$/L$_\odot$; 
and much higher ratios are unlikely.
This is within the range mentioned above, and gives us confidence that
the model is converging on a physically meaningful solution.

In summary, we conclude that the range of M$_{\rm BH}$ we
derive -- which we stress again is rather large due to the
uncertainties discussed in Sec.~\ref{input:kin} -- is reasonably
robust to the uncertainties in the input parameters.
The limits of the range are set not only by the Schwarzschild models
themselves but also by additional constraints on the mass to light
ratios which, through degeneracies between the parameters, affect 
the black hole mass.
The upper end is consistent with, although still less than, the other
mass estimates mentioned at the start of the Section; 
while the lower end is an order of magnitude smaller, 
suggesting that these techniques may tend to
overestimate the black hole mass.
Nevertheless, 
that the preferred black hole mass we derive via Schwarzschild
modelling is within a factor of 2--3 of the masses found by other
means and suggests that all methods are reasonable at least to this
level of accuracy.
With respect to the $M_{\rm BH}-\sigma_*$ relation, it is interesting
that in contrast to NGC\,3227 where we have found above that the stellar
dynamical mass is a 
factor of a few less than the relation predicts, the equivalent mass
for Cen\,A \citep{sil05} is a factor of several greater.
This suggests that, while the
$M_{\rm BH}-\sigma_*$ relation may be 
useful for order-of-magnitude $M_{\rm BH}$ estimates, the scatter for
active galaxies may be significantly larger than that for quiescent
galaxies.

\section{Molecular Gas and the Torus in NGC\,3227}
\label{sec:torus}

The 1-0\,S(1) morphology in Fig.~\ref{fig:fluxmaps} is remarkably
complex.
As has been noted by \cite{sch00} and \cite{bak00} on larger scales
for the cold molecular gas, we now find on smaller scales for the hot
molecular gas that 
the nucleus itself is not identified with the strongest emission.
The distribution is rather elongated at a position angle of
$-45^\circ$ (we note that there are inherent uncertainties and
difficulties due to the continuum subtraction associated with
the $100^\circ$ reported by \cite{qui99} and the
45--70$^\circ$ from \cite{fer99}).
It is tempting to interpret the emission in terms of a bar and arclets.
However, the kinematics and particularly the
surprisingly high velocity dispersion of 100--125\kms\ argue against it;
a more natural interpretation is in terms of a highly turbulent
medium, in which nevertheless the bulk of the gas follows uniform
rotation.
There do appear to be some non-rotational motions, but associated only
with weaker 1-0\,S(1) 
emission and restricted to the galaxy's minor axis.
This may indicate that some of the hot H$_2$ is influenced by outflows, or
that the emission from hot gas which we see here does not reflect well
the distribution of the cold gas.
Fig.~\ref{fig:velfield} shows that the velocity field of the brighter
emission -- corresponding to 2/3 of the flux in our field of view, and
hence also the bulk of the gas mass -- exhibits pure rotation.
These facts leads one towards the conclusion that most of the gas in
the nucleus probably exists in a thick rotating disk.

Fig.~\ref{fig:vsig} supports this interpretation.
It shows the rotation velocity $V_{\rm rot}$, corrected for
inclination, and the dispersion $\sigma$ for the molecular gas.
The data from the CO\,(2-1) data \citep{sch00} at larger
scales and the SINFONI data on smaller scales are fully consistent.
Also shown is the local $V_{\rm rot}/\sigma$ ratio (rather than the
global version which compares the maximum rotation velocity to the
central dispersion).
At $r\gtrsim1$\arcsec\ the ratio $V_{\rm rot}/\sigma$ clearly implies
that the gas lies in a thin disk. 
At smaller radii, the ratio decreases smoothly, indicating that the
gas distribution becomes geometrically thicker with respect to the
radial scales.

\subsection{Star Formation, Molecular Gas, and the Torus}
\label{sec:torusstars}

Above we argue that the gas lies in a thick disk,
and in Section~\ref{sec:star_prof} that this is also the most
likely distribution for the stars.
Although the morphologies appear different -- the stellar continuum is
more centrally concentrated and the isophotal position angle is closer
to north -- much of this arises from the bulge contibution to the
light.
Crucially, the kinematics for the gas and stars are remarkably
similar: the clear velocity gradient at PA -30 to -45$^\circ$, and the high
dispersion of more than 100\kms.
Because they exist on similar spatial scales of a few tens of parsecs
and exhibit similar kinematics, we conclude that the stars and gas
are physically mixed.

Scales of a few tens of parsecs are exactly those on which theoretical
models predict the obscuring torus should lie
\citep{pie92a,pie93,gra94,nen02,sch05}.
In addition, an important property of the torus is its ability to
collimate the UV radiation from the AGN to produce ionisation
cones whose apex lies close to the AGN, i.e. it is geometrically thick.
We have argued that our data show the gas and stellar distributions
are thick.
Our further conclusion is therefore that the gas and stars we see in
the nucleus of NGC\,3227 represent
much of what is understood by the term `obscuring torus'.

The idea of a star forming torus is not new, and has already been
modelled by \cite{wad02} and \cite{wad05} in response to the evidence
that about half of Seyfert~2 nuclei have a nuclear starburst.
However, from this work and that presented by \cite{dav04a,dav04b}, it
is now becoming clear also for type~1 Seyfert nuclei 
that star formation on scales associated with the torus is an
energetically important process.

If the gas associated with the 1-0\,S(1) emission were indeed part of the
obscuring torus, it would need to have a third 
crucial property: a high column density.
We address this issue in Section~\ref{sec:torusgas}, but first
consider below how the vertical height of the gas distribution might be
maintained.

\subsection{Supporting the Vertical Thickness of the Torus}
\label{sec:vert}

\begin{figure}
\epsscale{0.6}
\plotone{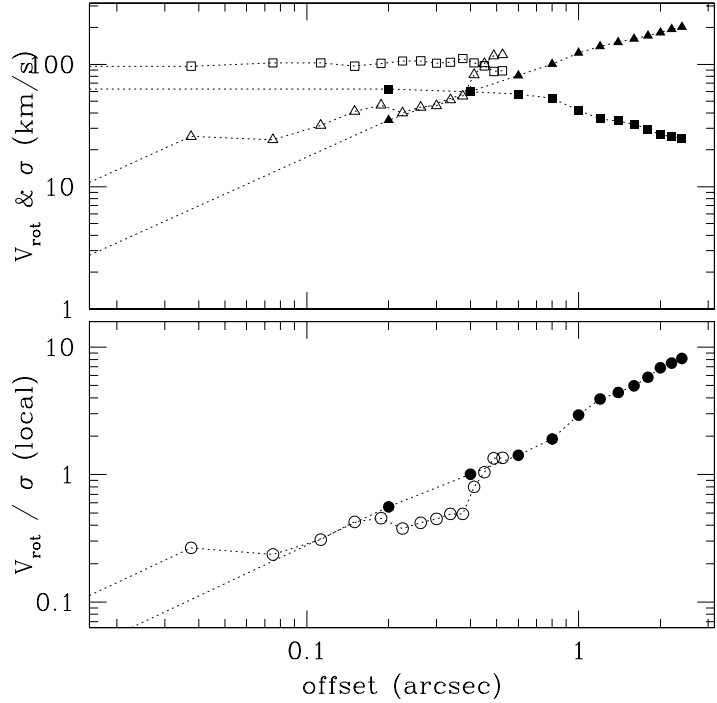}
\caption{Gas kinematics from the CO\,(2-1) data of \cite{sch00} and our
  SINFONI H$_2$ 1-0\,S(1) data (filled and empty points respectively).
Both are extracted along their kinematic major axis, with the two sides
  averaged.
Top: rotation velocity $V_{\rm rot}$ (triangles) and dispersion
  $\sigma$ (squares)
Bottom: the local $V_{\rm rot}/\sigma$ ratio increases smoothly with radius.
The differences between the two data sets are primarily due to the
  beam sizes (0.6\arcsec\ and 0.085\arcsec\ respectively).
}
\label{fig:vsig}
\epsscale{1.0}
\end{figure}

That the torus in NGC\,3227 is vertically extended is clear from
Fig~\ref{fig:vsig}: the
$\sim100$\kms\ velocity dispersion of the gas is larger than its
inclination corrected rotational velocity out to radii of
30--40\,pc.
Even the lower CO\,(2-1) dispersion of $\sim60$\kms\ is still
significat compared to the rotation velocity to these radii.
Thus the kinetic energy is dominated by random motions and implies a
more spheroidal structure rather than simple ordered rotation in a
thin disk.
Its constituent molecular clouds will undergo cooling via
collisions on short timescales, 
comparable to the orbital period of a few million years.
As a result energy must be constantly injected into the interstellar
medium in order to maintain the vertical structure.
The sources available to supply this energy are the 
AGN and the starburst, the latter via both radiation pressure and
mechanically through supernovae.
Such a situation may also provide, at least qualitatively, an answer to
why the star formation should have stopped while there is still a
large gas reservoir.

The effect of radiation pressure from the AGN itself has been explored
by \cite{pie92b}, and in their model depends primarily on the
Eddington ratio of the black hole and the clumpiness of the torus.
For the black hole mass, we adopt a value commensurate with our range
foudn in Section~\ref{sec:schwarz}, and those from X-ray variability
\citep{nik04} and reverberation mapping \cite{onk04}, that 
is M$_{\rm BH} = 2\times10^7$\,M$_\odot$.
For the AGN luminosity, we take half the bolometric luminosity since
the other half is supplied by the nuclear starburst, that is
L$_{\rm BH} = 10^{10}$L$_\odot$.
Since the Eddington luminosity 
L$_{\rm Edd} = 3.3\times10^4$M$_{\rm BH}$ in solar units, this yields
an Eddington ratio of L$_{\rm BH}$/L$_{\rm Edd} = 0.015$.
This value is an order of magnitude smaller than the ratio of 0.1 that
\cite{pie92b} found, was needed for a smooth torus to be thick and
static.
On the other hand, for tori consisting of large clumps, they found that in the
range $0.01 \lesssim L_{\rm Edd} \lesssim 0.1$ an equilibrium would
exist for thick tori with $1 \gtrsim a/h \gtrsim 0.1$ where $a$ is the
inner radius and $h$ is the full height in the body of the torus.
In the case of NGC\,3227 this could explain a thickness of up to a few
parsecs but not the ten to a few tens of parsecs implied by the velocity
dispersion, unless the AGN was much more active in the recent past.

The possibility that supernovae might heat the molecular torus
is treated specifically in the models of \cite{wad02}.
However, because the coupling from supernova energy to kinetic energy
of the ISM is rather weak, they used a very large supernova rate of 
$\nu_{\rm SN} \sim 1$\,yr$^{-1}$.
This is two orders of magnitude larger than the current rate in the
central 60\,pc of NGC\,3227, which is already close to the maximum for
our best fitting star formation history.
Hence, although they were able to reach scale heights for the torus of
10--20\,pc, it seems unlikely that this could work for NGC\,3227.

Recently, \cite{thomp05} have addressed the issue of whether radiation
pressure from the starburst itself can provide the vertical support.
For NGC\,3227 it is not clear that their optically thick model can be
applied, since we have argued in Section~\ref{sec:extinc} that the
extinction must be low enough 
that even in the near infrared the optical depth cannot exceed
$\tau_{\rm 2.2} = 2.6$ and is probably no more than $\tau_{\rm 2.2} = 1$.
In their optically thin limit -- where the optical depth to UV photons
is $\gtrsim$1 while to infrared photons from dust reprocessing it is
$\lesssim$1 -- they 
derive the star formation rate necessary to maintain marginal Toomre
stability ($Q \sim 1$), which for their model would give a scale
height of $\sim$10\,pc at a radius of 30\,pc.
If we assume a gas fraction of unity, and $\sigma\sim100$\kms\ as
measured for the 1-0\,S(1) line we find a star formation rate 
$\sim$20\,M$_\odot$\,yr$^{-1}$ would be needed.
This is two orders of magnitude greater than the current rate derived in
Section~\ref{sec:stars} but only a factor of 6--7 more than the peak
rate when the star formation was active.
There are two ways to reconcile this difference.
Using the 60\kms\ dispersion of the CO\,(2-1) data of
\cite{sch00}, one finds the star formation rate necessary to maintain
it is only 2.5\,M$_\odot$\,yr$^{-1}$.
One could then argue that the higher dispersion of the 1-0\,S(1) line
is due to the fact that this line traces only the hot gas, and
therefore will be strongly influenced by small scale turbulence.
Alternatively, one could argue that the gas fraction $f_{\rm g}$ is
less than unity:
for $f_{\rm g} = 0.5$, one derives a star formation rate of
5\,M$_\odot$\,yr$^{-1}$.

\begin{figure}
\epsscale{0.6}
\plotone{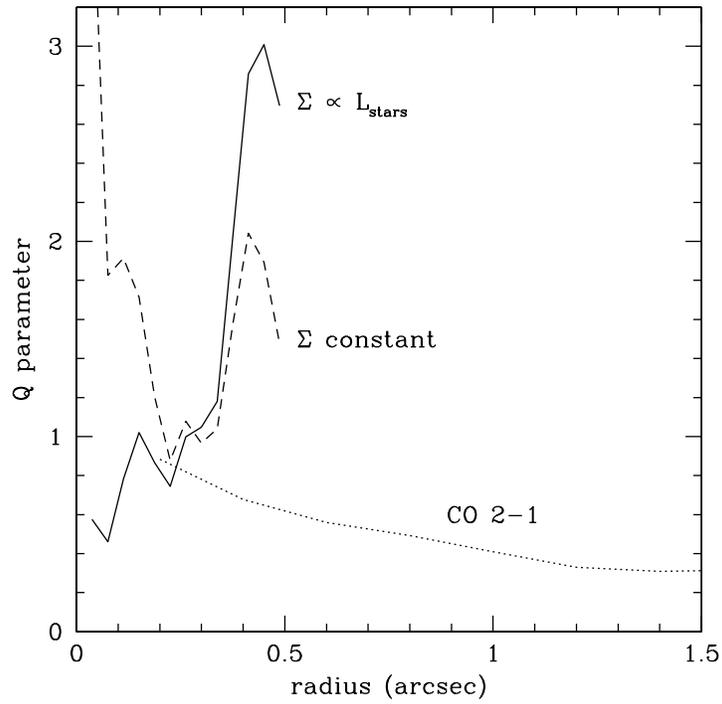}
\caption{Calculation of the Toomre $Q$ parameter, based on the
  kinematics in Fig.~\ref{fig:vsig} for the CO\,(2-1) data and our
  SINFONI 1-0\,S(1) data. For the former, the mass surface density
  $\Sigma$ is estimated under the assumption of Keplerian rotation in
  a thin disk; for the 1-0\,S(1) data we have considered the two extreme
  cases of $\Sigma$ being either constant or following the centrally
  concentrated stellar distribution.
}
\label{fig:toomre}
\epsscale{1.0}
\end{figure}

Either of these alternative estimates are reasonably consistent with
the peak rate derived from the observations.
However, they are both still far greater than the current rate.
It is perhaps conceivable that the molecular gas and star formation do
not reach an equilibrium state.
Instead, we speculate that what we are seeing is evidence for marginal
Toomre stability:
that once the gas cools, star formation may begin;
but that once radiation pressure from the young stars is
sufficiently high to heat the gas disk so that its turbulence
increases and it thickens, star formation can no longer proceed.
The molecular gas will then cool as the stars age, and sink back into
a thinner configuration, allowing the process to start again.
To investigate this in a more quantitative sense, 
we have calculated the Toomre $Q$
parameter, which defined in the usual way as 
\[
Q = \frac{\sigma \kappa}{\pi G \Sigma}
\]
where, using the data in Fig.~\ref{fig:vsig}, $\sigma$ is the velocity
dispersion and $\kappa$ is the epicyclic frequency.
The mass surface density $\Sigma$ we have estimated as follows.
For the CO\,(2-1) data it is derived from the velocity curve
under the approximation of Keplerian rotation in a thin disk.
For our SINFONI 1-0\,S(1) data, we have considered two cases: $\mu$ is
constant within the area considered; and $\Sigma$ follows the stellar
light distribution, being centrally concentrated.
For both of these we have assumed a total mass out to $r=0.5$\arcsec\
of $10^8$\,M$_\odot$ (see Section~\ref{sec:torusgas}).
The true distribution will lie somewhere between these two extremes.
The resulting estimates of $Q$ are shown in Fig.~\ref{fig:toomre}.
The difference in $Q$ derived from the 1-0\,S(1) and CO\,(2-1) data is a
direct result of the differing $\sigma$ apparent in
Fig.~\ref{fig:vsig} and to $\kappa$ which depends on the details of
the rotation curve.
What the figure shows is that as one approaches the nucleus from large
radii of 1--2\arcsec, $Q$ increases;
and within 0.5\arcsec of the nucleus, on average $Q > 1$.
Thus, the nuclear region does appear to be too dynamically hot to form
stars.
Based on the star formation timescale we derived earlier, we suggest
 that cycle of heating and cooling probably occurs on relatively short
 timescales of order 100\,Myr.
Thus time-dependent modelling may be necessary if the geometry of the
torus changes as the star formation or AGN fuelling pass through
successive active and quiescent phases.

\subsection{Mass and Column Density of the Gas}
\label{sec:torusgas}

The final thread in our analysis of whether the molecular gas in the
nucleus of NGC\,3227 can be identified with the obscuring torus
concerns its column density.
Our data trace only the hot $\gtrsim$1000\,K H$_2$ through the
1-0\,S(1) line, for which our flux of 1.1$\times10^{-17}$\,W\,m$^{-2}$
in a 0.8\arcsec\ aperture is
consistent with that in \cite{rod04} and \cite{qui99}.
Remarkably, it is possible to use this to make a reasonable estimate
of the total molecular gas mass.

That there exists a relation between molecular mass and 1-0\,S(1)
luminosity in actively star forming galaxies should not be surprising,
given that relations are already
known between 1-0\,S(1) and infrared luminosity $L_{\rm IR}$
\citep{gol97}, and between $L_{\rm IR}$ and CO luminosity $L_{\rm CO}$
which traces gas mass \citep{you91}.
By comparing the 1-0\,S(1) line luminosities of 17 luminous and
ultraluminous galaxies experiencing intense
star formation (some of which also host an AGN) to gas
masses derived from millimetre CO\,1-0 luminosities,
\cite{mul06} found that
\[ 
M_{\rm gas} [M_\odot] \, \sim \, 4000 \, L_{\rm 1-0S(1)} [L_\odot] 
\, \, \, \, {\rm (mean)}
\]
with a standard deviation of about a factor of 2.
The exception to this was NGC\,6240 which is known to be overluminous
in 1-0\,S(1) and for which the scaling factor is
by far the lowest, about an order of magnitude less than the
typical value.
The most likely reason for this is that the line does not originate in
star formation, but is due to cloud collisions resulting from the
extreme turbulence in the gas \citep{sug97,tac99}.
This particular case therefore provides us with a way to estimate a
conservative lower limit on the gas mass using
\[ 
M_{\rm gas} [M_\odot] \, \sim \, 430 \, L_{\rm 1-0S(1)} [L_\odot] \,
\, \, \, {\rm (NGC\,6240).}
\]
For NGC\,3227 this relation implies a gas mass in the central
0.8\arcsec\ (65\,pc) exceeding 4$\times10^7$\,$M_\odot$ and perhaps as
much as (2--8)$\times10^8$\,$M_\odot$.

We now compare this range of estimates to three other constraints on
the mass.
In Section~\ref{sec:stars} we found that the mass-to-light ratio of
the stars in this same nuclear region is 
$M_*/L_{\rm K} = 0.5 M_\odot/L_\odot$, with a young stellar 
mass of $2.0\times10^7$\,$M_\odot$.
On the other hand, the Schwarzschild modelling in
Section~\ref{sec:schwarz} implies that the most likely effective
mass-to-light ratio including gas, 
is $M_{\rm total}/L_{\rm K} = 1-5 M_\odot/L_\odot$.
And this result would imply a gas mass in the nuclear region of 
$M_{\rm gas} = 0.2-1.8\times10^8$\,$M_\odot$, and a most likely mass of
about half this.

The mass estimated by \cite{sch00} is rather less, but did not
include any correction to take into account the random motions
inferred from the large velocity dispersion.
We can make a rough estimate of the dynamical mass by including with
the Keplerian $M = V_{\rm rot}^2r/G$ the contribution of these random
motions so that $M \propto V_{\rm rot}^2+3\sigma^2$ (where the factor
3 comes from a simple comparison of the kinetic energies associated
with the two quantities).
Taking from the CO\,(2-1) data in Fig.~\ref{fig:vsig} that 
$V_{\rm rot}=50$\kms\ at 0.4\arcsec\ radius and $\sigma=60$\kms, we
find $M \sim 1.2\times10^8$\,M$_\odot$.

Finally, in Section~\ref{sec:vert} we argued that, if radiation
pressure does support the vertical thickness of the gas distribution,
then the gas fraction may be only $f_{\rm g} \sim 0.5$, implying equal
masses of stars and gas.
Thus $M_{\rm gas}$ could perhaps be as low as
$2.0\times10^7$\,$M_\odot$.

All the estimates above point to the same range of masses, and
so we can conservatively claim that the gas mass is most likely
2--$20\times10^7$\,$M_\odot$.
While there is an order of magnitude uncertainty, this still has
important implications.
Assuming the gas is uniformly distributed over the 0.8\arcsec\
aperture, the mean column density 
through the entire region would be 
$n_{\rm H} = {\rm (0.8}$--8)$\times10^{24}$\,cm$^{-2}$.
If the gas does represent the torus, then such a large column should
not be unexpected since there are many examples of type~2 AGN for
which hard X-ray observations indicate that the AGN itself is hidden
behind similar amounts of gas -- although it should be noted that the
X-ray absorption may occur on very small scales.
Two particularly well known cases are: NGC\,4945, for which the bright
nuclear emission below $\sim$10\,keV is completely blocked, implying
an absorbing column of $\sim5\times10^{24}$\,cm$^{-2}$
\citep{iwa93,don96};
and Circinus, for which the absorbing column was estimated to be
$\sim4\times10^{24}$\,cm$^{-2}$ based on the direct detection of
X-rays at energies greater than 10\,keV \citep{mat99}.

For a standard gas-to-dust ratio with 
$n_{\rm H} \, [cm^{-2}] \, = \, 1.5\times10^{21} \, \tau_{\rm V}$, 
the column density implies an optical depth in the visual of 
$\tau_{\rm V}>500$.
Since we are implicitly assuming that the stars are mixed with the gas
rather than hidden behind a foreground screen, we must again use the
mixed extinction model which we have used previously.
Hence we can compare directly the optical depth estimate here with the
maximum possible to the stars of $\tau_{\rm V} = 26$ from
Section~\ref{sec:stars}.
The two estimates are totally inconsistent with each other.
The solution to this contradiction lies in the gas, instead of
having a uniform distribution, being concentrated into clumps.
If the clumps are sufficiently compact then most lines of
sight will not be intersected -- and hence the stellar light will
suffer little extinction.
The degree of clumpiness need not be extreme: even local increases in
density of a factor of a few would be sufficient to reduce the covering
factor of the gas to half, consistent with the maximum stellar
extinction.

The gas density in these clumps will be high: the mean density over
the whole region for a
uniform distribution and assuming the gas is extended vertically as much
as it is laterally (i.e. by $\sim$60\,pc) is 
(4--40)$\times10^3$\,cm$^{-3}$.
This is already comparable to the typical
densities in parsec-scale cores of molecular clouds.
It is therefore no surprise that, as reported by \cite{rod04}, the
near infrared H$_2$ line ratios indicate that the hot gas is
thermalised.

In summary, we have used several independent mass estimates to put
limits on the gas mass within 30\,pc of the AGN.
This mass implies very significant column densities of gas, easily
sufficient to fulfil the third criterion of the obscuring torus
mentioned in Section~\ref{sec:torusstars}.
Thus, we can conclude that the gas we have observed via the 1-0\,S(1)
is likely to be associated with the molecular torus.
If so, then we have demonstrated that the gas in the torus is clumpy
in nature, and that the torus supports episodes of active star
formation.

\subsection{The Inner Edge of the Torus}
\label{sec:inneredge}

A curious result from the datacube is that the 2.1~and 2.3\micron\ continuum
centroids do not coincide.
Since the stellar continuum increases to shorter wavelengths, while
the non-stellar continuum (i.e. hot dust emission associated with the
AGN) increases to longer wavelengths, this can be most easily
explained if these two continua are offset from one another.
However, the scales involved are similar to those expected for
differential refraction between the two wavelengths, about
0.01\arcsec, and this makes the two phenomena difficult to
disentangle.

\begin{figure}
\epsscale{0.6}
\plotone{f20a.eps}
\epsscale{0.3}
\plotone{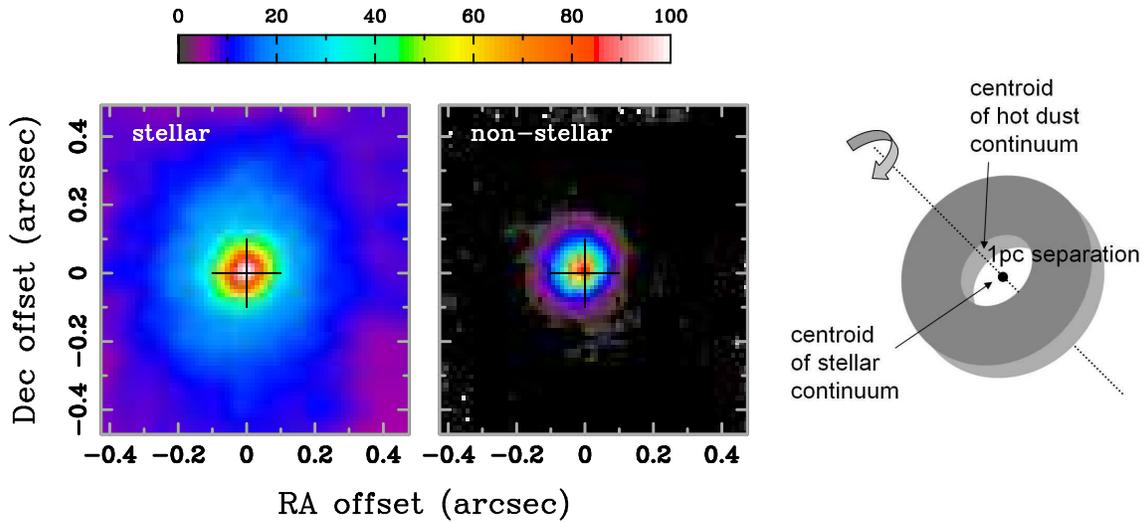}
\caption{Left and Centre panels: images from the SINFONI data cube
  showing the stellar and non-stellar continua, both at 2.3\micron.
The large cross on both panels indicates the centre of the stellar
  continuum.
The offset of the non-stellar continuum to the north east is easily
  seen.
Right panel: a cartoon of the canonical torus showing what might cause
  this offset.
North is up and east is left.
}
\label{fig:offset}
\epsscale{1.0}
\end{figure}

Instead, we can consider the separation of the continuum into its
stellar and non-stellar components via the CO bandheads longward of
2.3\micron. 
Because the two components are derived at the same wavelength, they are
not susceptible to differential refraction.
Yet the offset clearly remains, as shown in Fig.~\ref{fig:offset}: the
non-stellar continuum is shifted by about 1 pixel (0.0125\arcsec\ or
1\,pc, approximately 1/8 of the PSF size) to the north-east.
This effect is not due to a few pixels which happen by chance to be
brighter or fainter, but is a systematic global shift of the
entire feature;
and is significant with respect to the $0.05$\,pixel formal errors in
each axis from a Gaussian fit to the respective profiles.

Our interpretation is that we are seeing directly the inner edge of
the torus.
We have shown this schematically with the cartoon in the right panel of
Fig.~\ref{fig:offset}, where for simplicity we have drawn the
torus as a uniform ring with sharp edges; 
this representation should not be taken too literally.
If the torus is oriented in the same way as the larger
(i.e. $\gtrsim$100\,pc) scales, it will be inclined with its major
axis on a line from south-east to north-west.
As a result, the inner edge will only be visible on the far side; the near
side will be hidden by the outer layers, since the torus is presumably
optically thick.
Regardless of whether the torus is modelled as a uniform or clumpy
medium, its inner boundary is always expected to be a radius of
$\sim$1\,pc \cite[e.g.][]{sch05}.
This distance is set by the sublimation temperature of the dust
grains, since they cannot exist any closer to the AGN.
Depending on the thickness and inclination of the torus, one then
would expect the centroid of the observable emission from the hottest
dust grains (i.e. those at $\sim1000$\,K, heated by the AGN rather
than by stars)
to lie at a projected distance of $\sim$1\,pc from the AGN itself.
In contrast, since star formation occurs throughout the torus, the
centroid of the stellar continuum is centered on the AGN.
Hence one would expect to find, and indeed Fig.~\ref{fig:offset}
shows, an offset between the stellar and
hot dust continua.

\section{Conclusions}
\label{sec:conc}

We have presented new near infrared integral field spectroscopic data
at 70\kms\ FWHM spectral resolution of the Seyfert~1 nucleus in
NGC\,3227, making use of adaptive optics to reach a spatial resolution
of 0.085\arcsec\ (7\,pc).
In this paper we have addressed the issues of star formation and
molecular gas around the
AGN and their relation to the canonical obscuring torus;
and we have analysed the stellar kinematics using Schwarzschild models
to determine the black hole mass. 
Our main conclusions are:

\begin{itemize}

\item
The nuclear star forming region around the Seyfert~1 nucleus is
spatially resolved, on scales of a few parsecs to a few tens of
parsecs.
The most recent episode of intense star formation began $\sim$40\,Myr
ago but has now ceased.
Within 30\,pc of the AGN this starburst still accounts for 20--60\% of
the galaxy's bolometric luminosity.
Despite showing evidence for moderate rotation, the stars' kinetic
energy is dominated by random motions indicating that they lie in a
thick disk.

\item
Schwarzschild modelling of the stellar kinematics lead to a black hole
mass in the range M$_{\rm BH}=7 \times 10^6-2\times 10^7$\,M$_\odot$.
The upper end is consistent with (although still less than) previous
estimates made using other techniques.
The large range arises through a degeneracy in whether mass is
attributed to the black hole or the stars and gas, which can be
resolved with better kinematic line profiles.

\item
The gas in the central 80\,pc of NGC\,3227, exhibits several critical
properties that are expected of a molecular obscuring torus: its
spatial extent is a few tens of 
parsecs, it is geometrically thick, and it has column density of order 
$n_{\rm H} = 10^{24}$--$10^{25}$\,cm$^{-2}$.
This argues that the gas we have observed is the torus.
Moreover, based on the similarity of their spatial extents and their
kinematics, it is likely that the gas and stars are physically mixed.
Thus the torus also supports episodes of active star formation.

\item
It seems unlikely that the current level of AGN or star forming
activity can inject sufficient energy into the ISM to maintain the
vertical thickness of the torus.
However, this was possible when the star formation rate was at its
peak value.
We speculate that the torus may heat and subsequently cool, changing
its vertical profile, as the star
formation and AGN go through active and quiescent phases.

\end{itemize}


\acknowledgments

The authors are grateful to the staff at the Paranal Observatory for
their support during the observations, and for the entire SINFONI team
at MPE and ESO.
They are also indebted to Carole Mundell for kindly finding and
re-reducing the radio continuum data.
Finally, they thank the referee for comments which helped improve
the manuscript.



\clearpage


\begin{deluxetable}{lcl}
\tablecaption{Measured fluxes and flux densities for NGC\,3227\label{tab:flux}}
\tablewidth{0pt}
\tablehead{
\colhead{Item} &
\colhead{Measurement\tablenotemark{a}} &
\colhead{unit} \\
}

\startdata

H$_2$ 1-0\,S(1)          & 10.9\tablenotemark{b}
 & $10^{-18}$\,W\,m$^{-2}$ \\

narrow Br$\gamma$        & \phm{0}4.6\tablenotemark{b,c}
 & $10^{-18}$\,W\,m$^{-2}$ \\

broad Br$\gamma$         & \phm{0}42\tablenotemark{b}\phm{.0} 
 & $10^{-18}$\,W\,m$^{-2}$ \\

[Ca\,{\sc viii}]         & \phm{0}1.5\tablenotemark{b}
 & $10^{-18}$\,W\,m$^{-2}$ \\

total K-band continuum   & \phm{0}32\phm{$^a$.0} 
 & $10^{-15}$\,W\,m$^{-2}$\,\micron$^{-1}$ \\

$W_{\rm CO}$             & \phm{0}3.6\tablenotemark{d}
 & \AA \\

young stellar K-band continuum & \phm{0}7.0\tablenotemark{e} 
 & $10^{-15}$\,W\,m$^{-2}$\,\micron$^{-1}$ \\

old stellar K-band continuum & \phm{0}4.5\tablenotemark{e} 
 & $10^{-15}$\,W\,m$^{-2}$\,\micron$^{-1}$ \\

\enddata

\tablenotetext{a}{\,All measurements are given for a 0.8\arcsec\
  aperture centered on the continuum peak.}

\tablenotetext{b}{\,Uncertainties are approximately
  $0.3\times10^{-18}$\,W\,m$^{-2}$ (3 times larger for the broad
  Br$\gamma$), and are dominated by calibration 
  and from the stellar continuum features.}

\tablenotetext{c}{\,We estimate that 25\% originates in the narrow
  line region and 75\% from star formation, yielding the number of
  ionising photons from young stars to be 
  $\log{Q_{\rm Lyc}} = 51.93$.}

\tablenotetext{d}{\,The dilution at 2.3\micron\ implied by
  $W_{\rm CO}$ is greater than the mean dilution over the K-band which
  is centered at 2.18\micron.}

\tablenotetext{e}{\,Assuming a dilution over the K-band of 2.8; the
  uncertainty in this corresponds to no more than 20\% uncertainty in
  stellar the continuum flux density. The division between young and
  old populations is based on the bulge and excess continua as shown
  in Fig.~\ref{fig:prof}.}

\end{deluxetable}

\begin{deluxetable}{lcl}
\tablecaption{Derived properties for the central 0.8\arcsec\ of
  NGC\,3227\label{tab:derived}}
\tablewidth{0pt}
\tablehead{
\colhead{Item} &
\colhead{Measurement\tablenotemark{a}} &
\colhead{unit} \\
}

\startdata

$\log{L_K/L_\odot}$ (stellar) & 7.8 & \\
$M_{\rm gas}$                 & 2--$20\times10^7$ & M$_\odot$ 
\smallskip\\

\multicolumn{3}{l}{\it young stellar component:} \\

$\log{L_K/L_\odot}$           & 7.6  & \\
$\log{L_{\rm bol}/L_\odot}$   & 9.5  & \\
initial star formation rate   & 3    & M$_\odot$\,yr$^{-1}$ \\
current star formation rate   & 0.05 & M$_\odot$\,yr$^{-1}$ \\
age                           & 40   & Myr \\
e-folding decay time          & 10   & Myr \\
$M_*/L_K$                     & 0.5  & M$_\odot$/L$_\odot$ \\
$M_*$                         & $2\times10^7$ & M$_\odot$ \\

\enddata

\end{deluxetable}


\clearpage

\clearpage

\begin{figure}
\epsscale{0.4}
\plotone{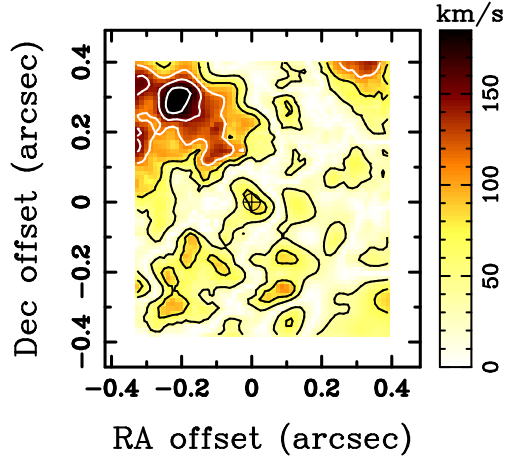}
\caption{Absolute difference between the H$_2$ 1-0\,S(1) and stellar velocity
  fields, with contours drawn at intervals of 25\kms.
The only major difference between the two is to the north-east, out
  along the minor axis.
}
\label{fig:veldiff}
\epsscale{1.0}
\end{figure}


\end{document}